\documentclass[12pt,a4paper,dvips]{article}
\usepackage{amssymb,amsmath,graphicx}
\usepackage{cite,mcite}
\usepackage{a4p}
\usepackage{ifthen}
\usepackage{l3_title}
\usepackage{Lep}
\usepackage{higgs_input}
\Lep{2}
\journalname{Phys. Lett. B}
\date{October 24, 2003}
\preprint{2003-069}       

\newlength{\capindent}
\newlength{\capwidth}
\newlength{\figwidth}
\newcommand{\icaption}[2][!*!,!]{\hspace*{\capindent}%
  \begin{minipage}{\capwidth}
    \ifthenelse{\equal{#1}{!*!,!}}%
      {\caption{#2}}%
      {\caption[#1]{#2}}
  \end{minipage}}

\newcommand{\mh}{m_\mathrm{h}}

\newcommand{\mA}{m_\mathrm{A}}
\newcommand{\mZ}{m_\mathrm{Z}}
\newcommand{\ee}{$\mathrm{e^+e^-}$}
\newcommand{\ghZZ}{g_{\mathrm{hZZ}}}
\newcommand{\ghAZ}{g_{\mathrm{hAZ}}}
\newcommand{\gSM}{g_{\mathrm{HZZ}}^{\rm SM}}
\def\ra{\rightarrow}

\begin{document}

\begin{titlepage}

\title{ 
  Flavour Independent Search for\\
  Neutral Higgs Bosons at LEP
}

\author{The L3 Collaboration}

%%%%%%%%%%%%%%%%%%%%%%%%%%%%%%%%%%%%%%%%%%%%%%%%%%%%%%%%%%%%%%
\begin{abstract}
%%%%%%%%%%%%%%%%%%%%%%%%%%%%%%%%%%%%%%%%%%%%%%%%%%%%%%%%%%%%%%

A flavour independent search for the CP-even and CP-odd neutral Higgs
bosons \h and \A is performed in 624 pb$^{-1}$ of data collected with
the L3 detector at LEP at centre-of-mass energies between 189 and
$209\GeV$.  Higgs boson production through the $\rm \epem\ra Z h$ and
the $\rm \epem\ra\h\A$ processes is considered and decays of the Higgs
bosons into hadrons are studied. No significant signal is observed and
95\% confidence level limits on the hZZ and hAZ couplings are derived
as a function of the Higgs boson masses. Assuming the Standard Model
cross section for the Higgs-strahlung process and a 100\% branching
fraction into hadrons, a 95\% confidence level lower limit on the mass
of the Higgs boson is set at 110.3\GeV{}.
\end{abstract}

\submitted

\end{titlepage}

%%%%%%%%%%%%%%%%%%%%%%%%%%%%%%%%%%%%%%%%%%%%%%%%%%%%%%%%%%%%%%
\section{Introduction}
%%%%%%%%%%%%%%%%%%%%%%%%%%%%%%%%%%%%%%%%%%%%%%%%%%%%%%%%%%%%%%

One of the goals of the LEP program is the search for Higgs bosons,
the particles postulated by the Standard Model of the electroweak
interactions~\cite{sm_1}, and some of its extensions, to
explain the mechanism~\cite{higgs_1} which gives the
elementary particles their observed masses.

At the centre-of-mass energies, $\sqrt{s}$, at which the LEP \ee
collider was operated, the Standard Model Higgs boson, H, is predicted
to decay dominantly into b quarks.  For a large part of the parameter
space of the Minimal Supersymmetric Standard Model
(MSSM)~\cite{mssm_1}, decays of neutral Higgs bosons into b quarks are
also predicted to be dominant. Experimental searches for the Higgs
bosons predicted both in the Standard Model and in the MSSM exploit
this feature through sophisticated flavour tagging techniques. No
significant signal was found at LEP either for the Standard Model
Higgs boson~\cite{l3_sm_paper,aleph_mssm,opal_sm,lnq} or for neutral
Higgs bosons of the MSSM~\cite{l3_mssm_paper,aleph_mssm,opal_mssm}.

In some extensions of the Standard Model, decays of the Higgs bosons
into $\mathrm{b\bar{b}}$ pairs are strongly suppressed to the benefit
of other decay modes such as $\mathrm{c\bar{c}}$, gg or
$\mathrm{\tau^+\tau^-}$. For instance, this occurs for specific
parameters of the two Higgs doublet model~\cite{higgs_hunters} or the
MSSM~\cite{mssm_benchmark}, as well as for some composite
models~\cite{composite}.  It is hence important to investigate such
scenarios with dedicated experimental analyses in which the
information about the flavour of the Higgs boson decay products is not
used, reducing the model dependence of the conventional Higgs
searches.

This Letter describes the search for hadronic decays of 
the light CP-even Higgs boson, \h, and of the CP-odd Higgs boson, \A,
using data collected by the L3 detector~\cite{l3_det} 
at LEP. Production of a \h boson 
in association with a Z boson, Higgs-strahlung, and
pair-production of the \h and \A bosons, are considered: 
\begin{displaymath}
\mathrm{
e^+e^-\rightarrow hZ, 
}\,\,\,\,\,\,\,\,
\mathrm{
e^+e^-\rightarrow hA. 
}
\end{displaymath}
The tree level cross sections of these processes are related to the  
cross section of the Standard Model Higgs boson production through the
Higgs-strahlung process, $\sigma_{\rm HZ}^{\rm SM}$, 
as~\cite{higgs_hunters}:
\begin{displaymath}
\mathrm{
\sigma_{hZ}=\xi^2\sigma_{HZ}^{SM},
}\,\,\,\,\,\,\,\,
\mathrm{
\sigma_{hA}=\eta^2\tilde{\lambda}\sigma_{HZ}^{SM},
}
\end{displaymath}
where $\mathrm{\tilde{\lambda}}$ is a p-wave 
suppression factor, which depends on $\sqrt{s}$ and on the Higgs boson
masses, $\mh$ and $\mA$. The hZZ and hAZ couplings relative to the HZZ
coupling of the Standard Model are defined as $\xi=\ghZZ/\gSM$ and
$\eta=\ghAZ/\gSM$. In the following, these couplings are not fixed to
any prediction but rather considered as free parameters, reducing the
model dependence of the analysis.

%%%%%%%%%%%%%%%%%%%%%%%%%%%%%%%%%%%%%%%%%%%%%%%%%%%%%%%%%%%%%%
\section{Data and Monte Carlo samples}
%%%%%%%%%%%%%%%%%%%%%%%%%%%%%%%%%%%%%%%%%%%%%%%%%%%%%%%%%%%%%%

An integrated luminosity of 624 pb$^{-1}$ of data, collected at
$\sqrt{s}=189-209\GeV$, is analysed. The data are grouped into several
subsamples according to their $\sqrt{s}$ value, as listed in
Table~\ref{tab:data}.

The cross section of the Higgs-strahlung process in the Standard Model
is calculated using the HZHA generator~\cite{hzha}. For efficiency
studies, Monte Carlo samples are generated using PYTHIA~\cite{pythia}
for the two production mechanisms and for each of the decay modes
$\h\rightarrow\mathrm{b\bar{b}}$, $\mathrm{c\bar{c}}$ and
$\mathrm{gg}$, $\A\rightarrow\mathrm{b\bar{b}}$, $\mathrm{c\bar{c}}$
and $\mathrm{gg}$. Several Higgs mass hypotheses are considered and
2000 events are generated in each case.  For the $\epem\rightarrow\hZ$
process, $\mh$ ranges in steps of 10\GeV from 60 to 100\GeV, and in
steps of 1\GeV from 100 to 120\GeV. For the $\epem\rightarrow\hA$
process, $\mh$ and $\mA$ range from 40 to 110\GeV in steps of 10\GeV.

For background studies, the following Monte Carlo programs are used:
KK2f~\cite{kk2f} for \epemtoqqg and \epemtotautau, PYTHIA for
\epemtoZZ and $\epem\rightarrow\Z\epem$ and YFSWW~\cite{YFSWW} for
\epemtoWW. EXCALIBUR~\cite{excalibur} is used for four-fermion final
states not covered by these generators.  Hadron production in
two-photon interactions is simulated with PYTHIA and
PHOJET~\cite{phojet}.  The number of simulated events for the most
important background channels is more than 100 times the number of
expected events.

The L3 detector response is simulated using the GEANT
program~\cite{geant}, which models the effects of energy loss,
multiple scattering and showering in the detector. The GHEISHA
program~\cite{gheisha} is used to simulate hadronic interactions. Time
dependent detector inefficiencies, monitored during data taking, are
also taken into account.

%%%%%%%%%%%%%%%%%%%%%%%%%%%%%%%%%%%%%%%%%%%%%%%%%%%%%%%%%%%%%%
\section{Analysis procedures}
%%%%%%%%%%%%%%%%%%%%%%%%%%%%%%%%%%%%%%%%%%%%%%%%%%%%%%%%%%%%%%

Three different decay modes are considered for the \h and \A bosons:
$\mathrm{b\bar{b}}$, $\mathrm{c\bar{c}}$ and $\mathrm{gg}$.
Table~\ref{tab:signatures} summarises the different signal signatures
and the investigated topologies. Three topologies cover the possible
final states of the $\epem\ra\hZ$ process. They correspond to the
decay of the Z boson into hadrons, neutrinos or charged leptons,
associated to the hadrons from the \h decay. They give rise to events
with four hadronic jets, two hadronic jets and missing energy and two
hadronic jets and two charged leptons, respectively. A single topology
consisting of four hadronic jets, covers all final states of the
$\epem\ra\hA$ process.

Analyses in all channels proceed from a preselection of high
multiplicity hadronic events which suppresses copious backgrounds from
two-photon interactions, lepton-pair production and pair-production of
gauge bosons which decay into leptons. A selection based on kinematic
cuts, neural networks or likelihoods is then applied to further
discriminate the signal from the background.  Finally, discriminant
variables which depend on the Higgs mass hypothesis are built to
separate signal and background. Their distributions are studied to
test the presence of a signal and to probe the $\xi$ and $\eta$
couplings as a function of $\mh$ and $\mA$. Events are ordered as a
function of the signal over background ratio and only events with this
ratio greater than 0.05 are retained.

%%%%%%%%%%%%%%%%%%%%%%%%%%%%%%%%%%%%%%%%%%%%%%%%%%%%%%%%%%%%%%
\section{Search for \boldmath$\rm e^+e^-\rightarrow hZ$}
%%%%%%%%%%%%%%%%%%%%%%%%%%%%%%%%%%%%%%%%%%%%%%%%%%%%%%%%%%%%%%

The three analyses used in the search for $\rm e^+e^-\rightarrow hZ$
are similar to those used in the search for the Standard Model Higgs
boson~\cite{l3_sm_paper}, with the exception that no b quark
identification is used.

%%%%%%%%%%%%%%%%%%%%%%%%%%%%%%%%%%%%%%%%%%%%%%%%%%%%%%%%%%%%%%
\subsection{Four jets}
%%%%%%%%%%%%%%%%%%%%%%%%%%%%%%%%%%%%%%%%%%%%%%%%%%%%%%%%%%%%%%

If both the h and the Z bosons decay into hadrons, the
signature is four hadronic jets. The invariant mass of two of
them has to be compatible with the mass of
the Z boson, $\mZ$.  The dominant background comes from hadronic
decays of pair-produced gauge bosons and from the
$\mathrm{e^+e^-\rightarrow q\bar{q}(\gamma)}$ process.

After a preselection of high multiplicity events~\cite{l3_sm_paper},
events are resolved into four jets using the DURHAM
algorithm~\cite{DURHAM} and a kinematic fit imposing four-momentum
conservation is performed. A likelihood, $L_{\mathrm{hZ}}$, is
built~\cite{l3_sm_paper} from the following variables:

\begin{itemize}
\item the maximum energy difference between any two jets, 
\item the minimum jet energy, 
\item the parameter $y_{34}$ of the DURHAM algorithm for which the 
    event is resolved from  three
    into  four jets,  
\item the minimum opening angle between any two jets, 
\item the event sphericity, 
\item the  absolute value of the cosine of the polar angle,
$\Theta_{\rm 2B}$, for the di-jet system most compatible with the
production of a pair of gauge  bosons,
\item the mass from a kinematic fit
imposing four-momentum conservation and 
equal di-jet masses, ${m_{\rm 5C}}$, 
\item the maximal triple-jet boost, $\gamma_\mathrm{triple}$,
    defined as the maximum three-jet boost obtained
    from the four possibilities to construct a 
    one-jet against three-jet configuration in a four-jet event.
\end{itemize}

\begin{figure}
\begin{center}
\begin{tabular}{cc}
\hspace{-7mm}
\includegraphics*[width=0.5\textwidth]{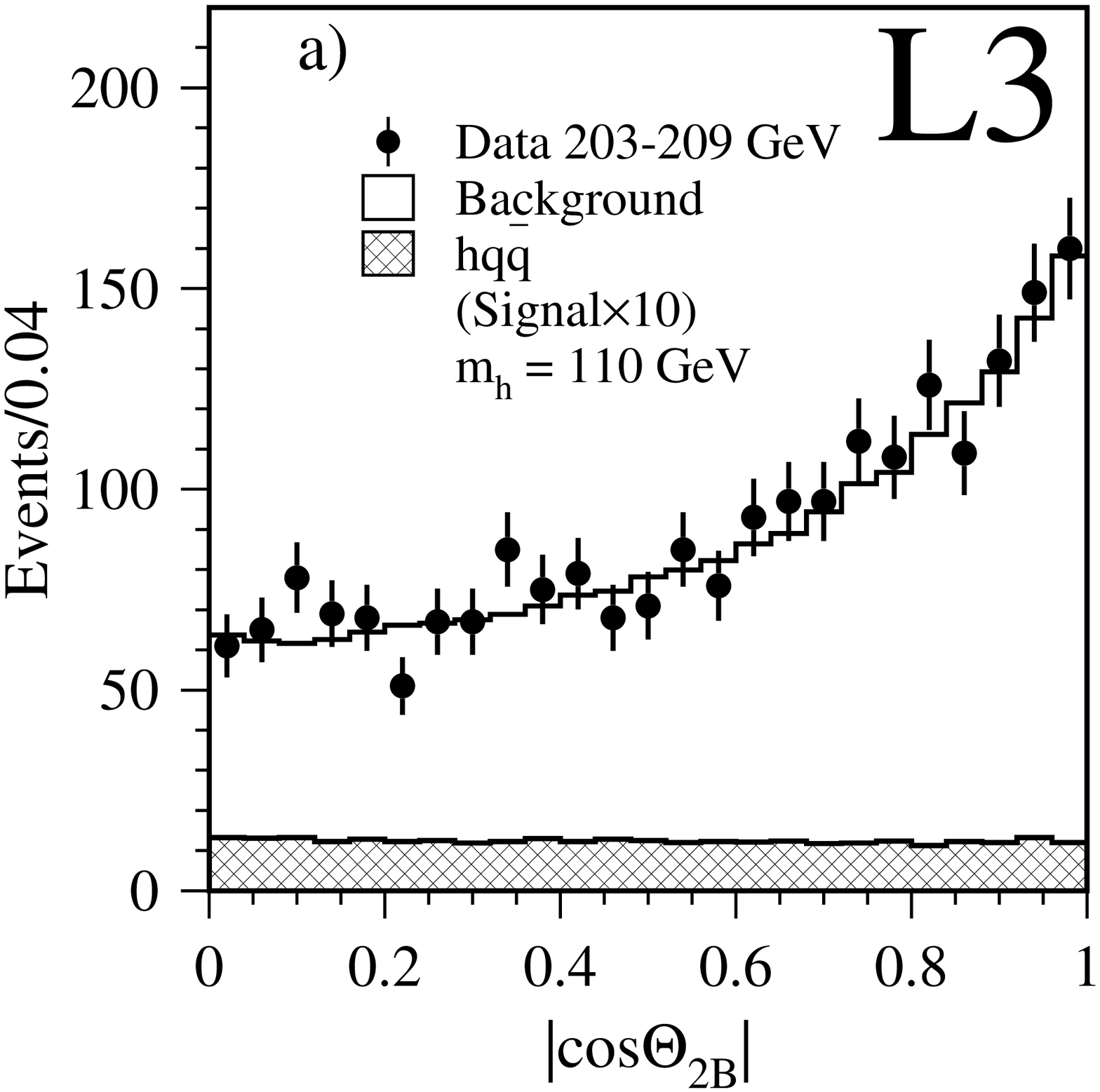} &
\hspace{-0mm}
\includegraphics*[width=0.5\textwidth]{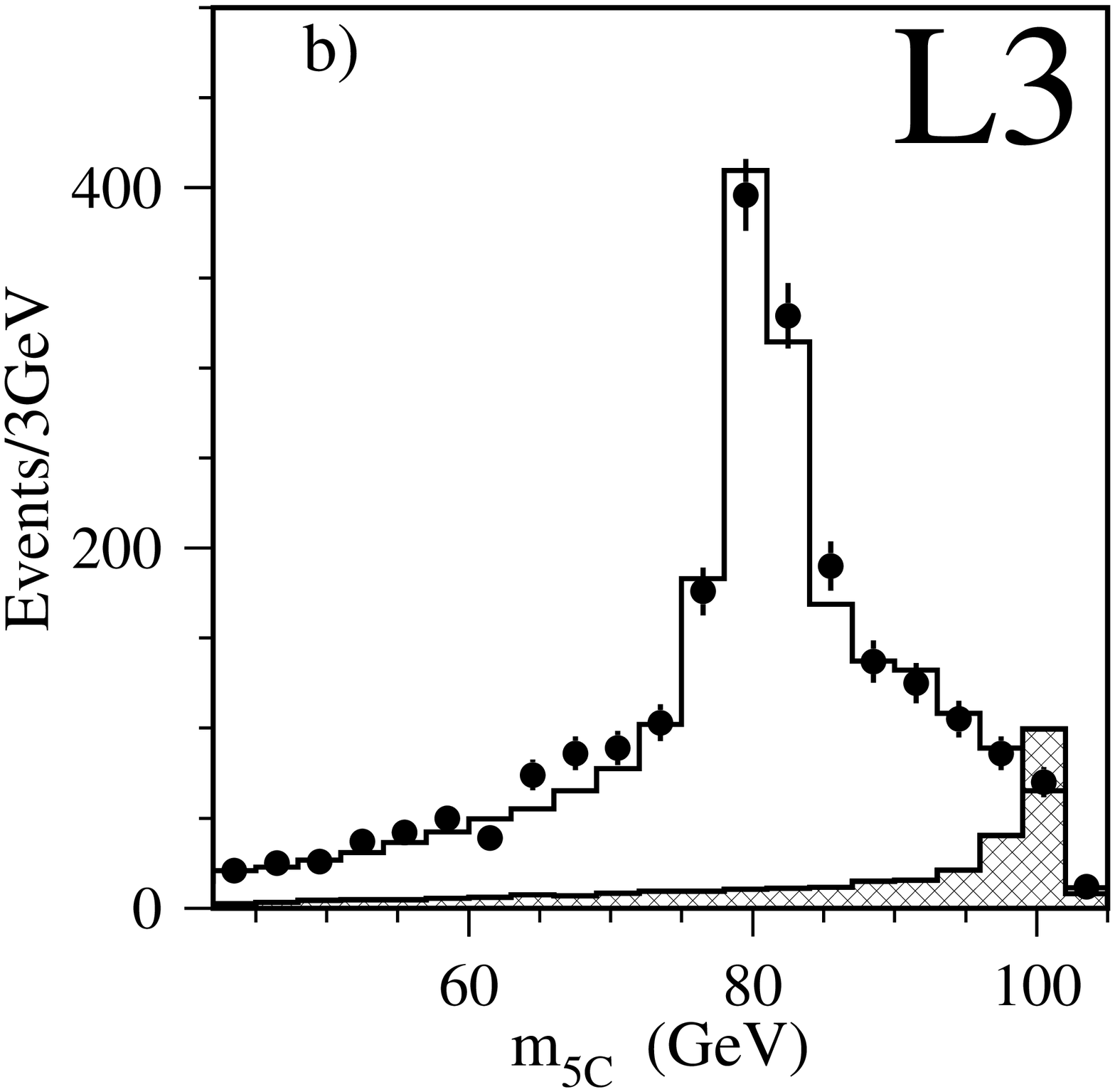} \\
\hspace{-7mm}
\includegraphics*[width=0.5\textwidth]{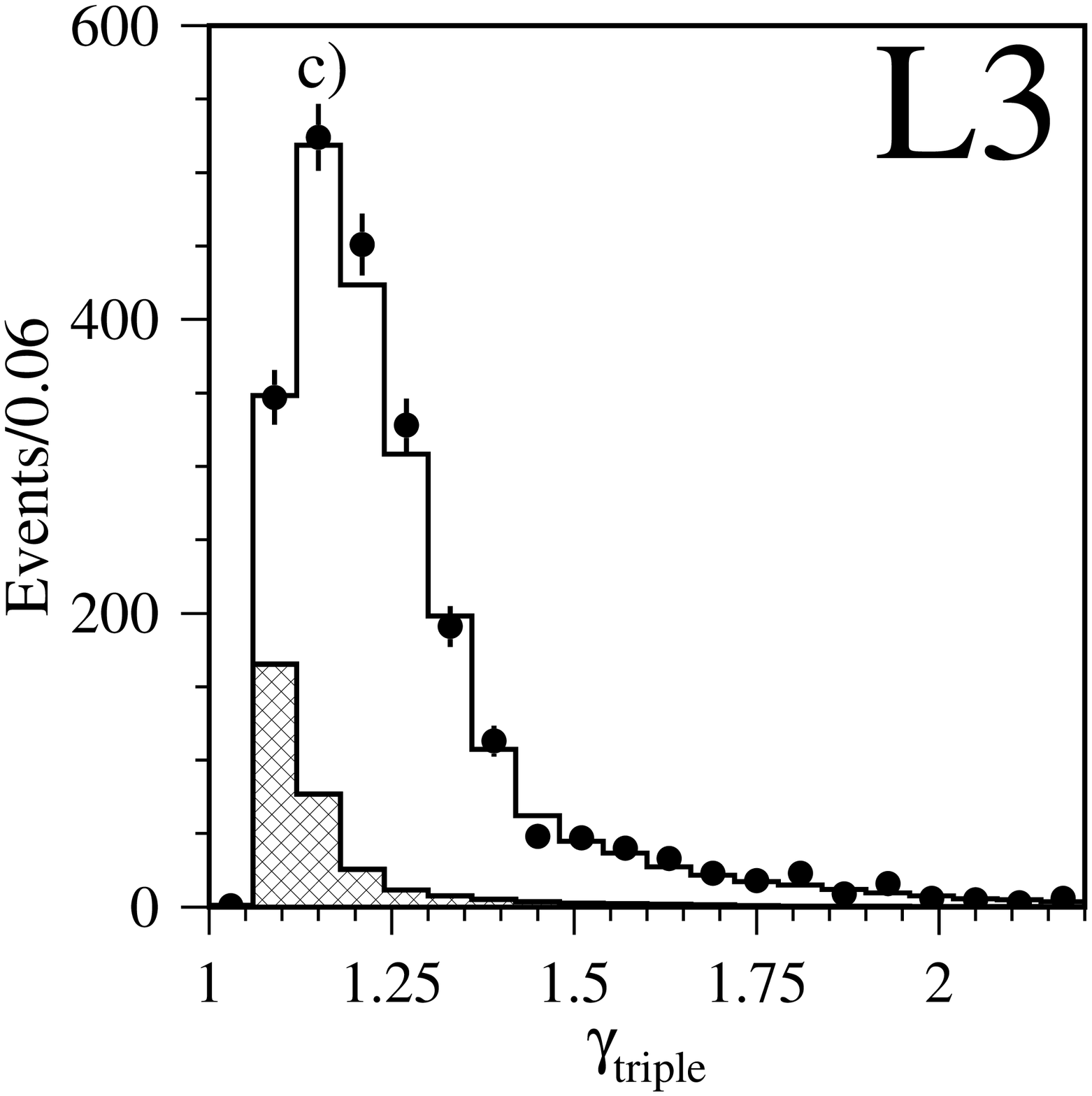} &
\hspace{-0mm}
\includegraphics*[width=0.5\textwidth]{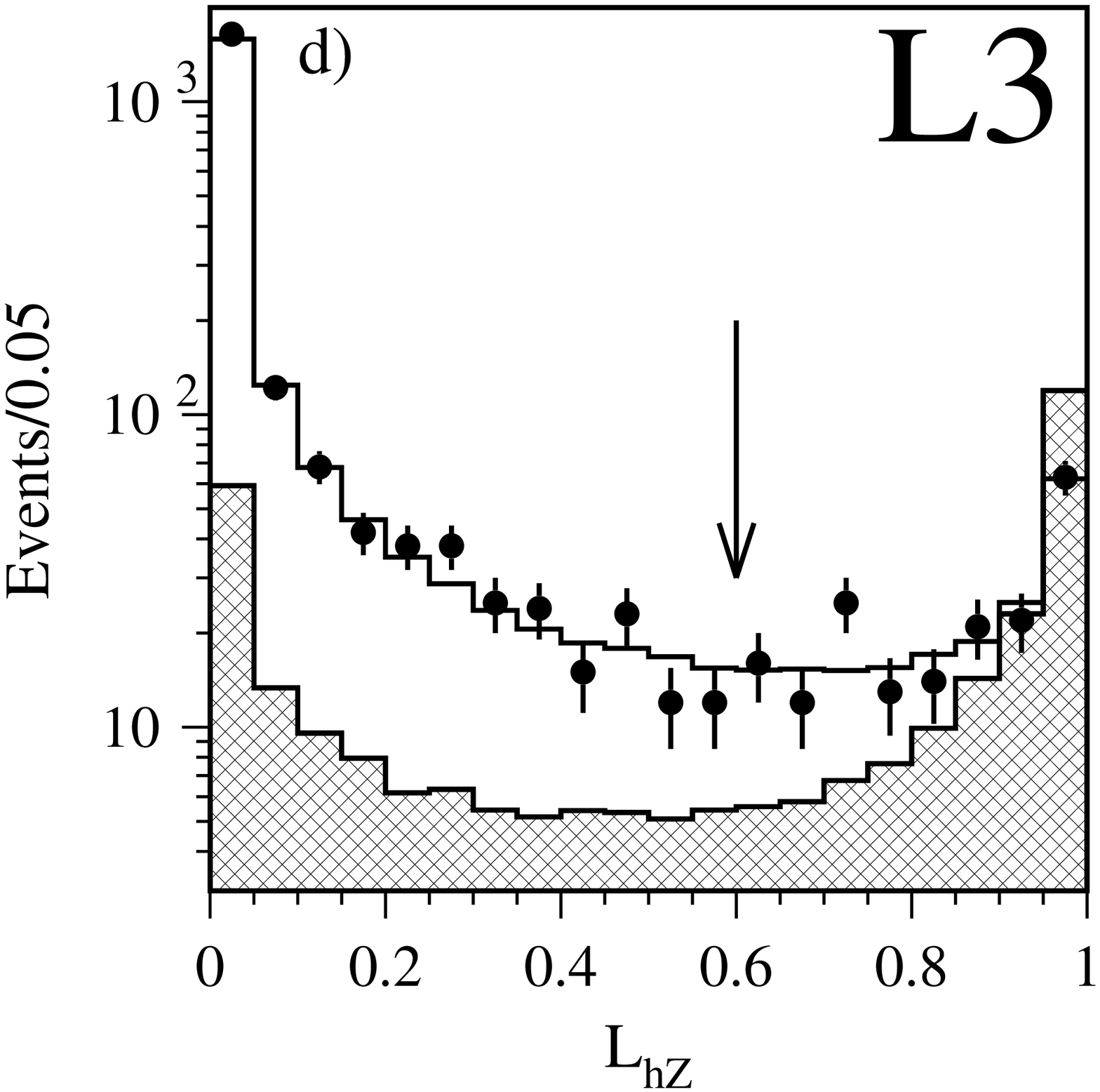} \\
\end{tabular}
        \caption[]{\label{fig:hqq}
        Distributions for the $\rm e^+e^-\rightarrow hZ$ search in the
        four-jet final state of  a) $|\cos{\Theta_\mathrm{2B}}|$,
        b) $m_\mathrm{5C}$,
        c) $\gamma_\mathrm{triple}$,
        d) $L_\mathrm{hZ}$.
        The points indicate data collected at $\sqrt{s}$ $>$ 203\GeV,
        the open histograms represent the expected background
        and the hatched histograms stand for a $\mh$ = 110\GeV signal 
        expected for $\mathrm{\xi^2\times B(h\ra hadrons)}$ = 1, 
        multiplied by a factor of 10. The arrow in d) indicates
        the position of the cut.}
\end{center}
\end{figure}

Figure~\ref{fig:hqq} shows the distributions of $|\cos{\Theta_{\rm
2B}}|$, $m_\mathrm{5C}$, $\gamma_\mathrm{triple}$ and
$L_\mathrm{hZ}$ for data collected at $\sqrt{s}$ $>$ 203\GeV, the
expected background and a signal with $\mh = 110\GeV$.  Events are
retained for which the value of $L_{\mathrm{hZ}}$ exceeds a threshold,
around 0.6, optimised separately for each $\sqrt{s}$ and $\mh$
hypothesis.

\begin{figure}
\begin{center}
\begin{tabular}{cc}
\hspace{-7mm}
\includegraphics*[width=0.47\textwidth]{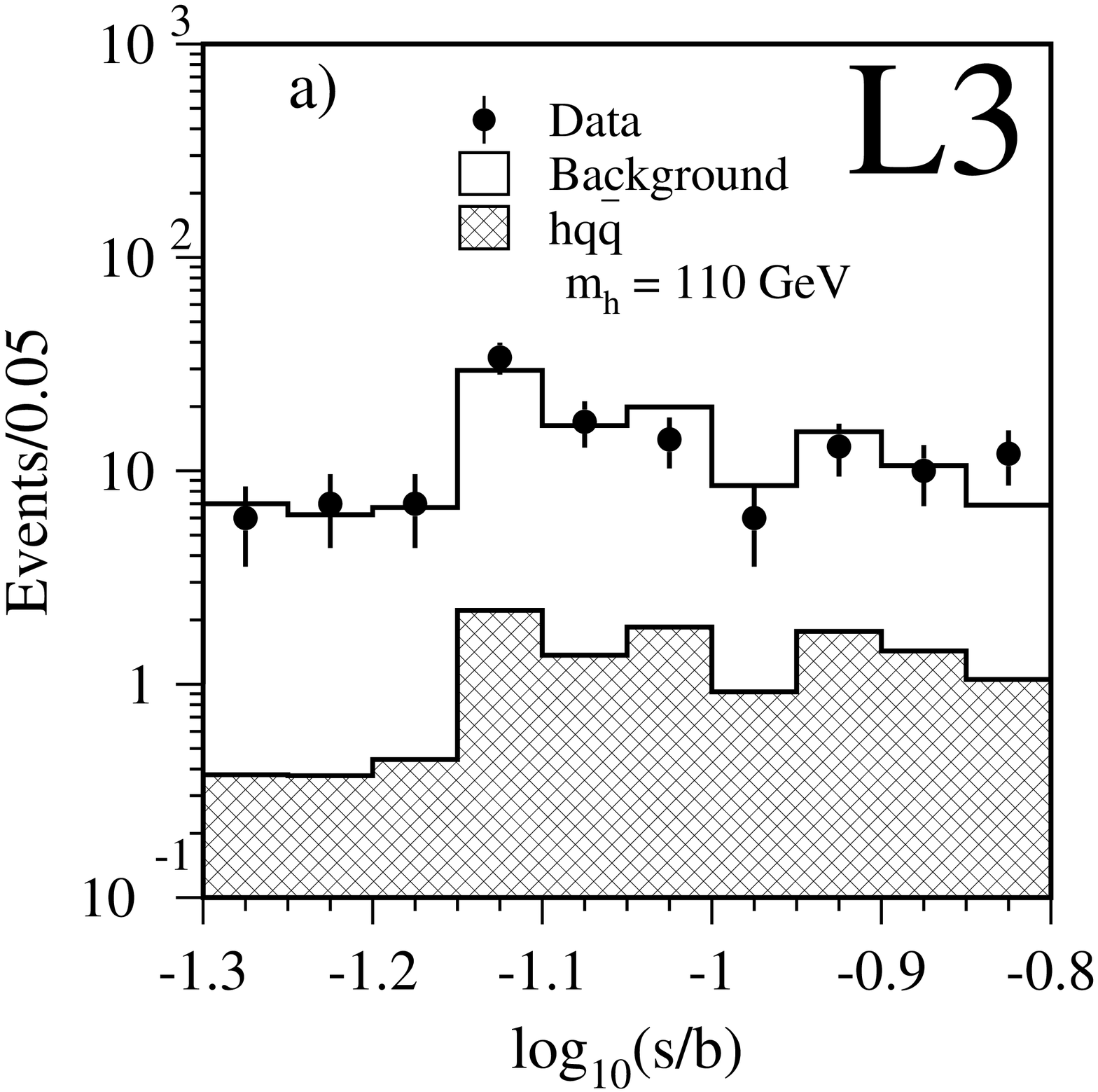} &
\hspace{-0mm}
\includegraphics*[width=0.47\textwidth]{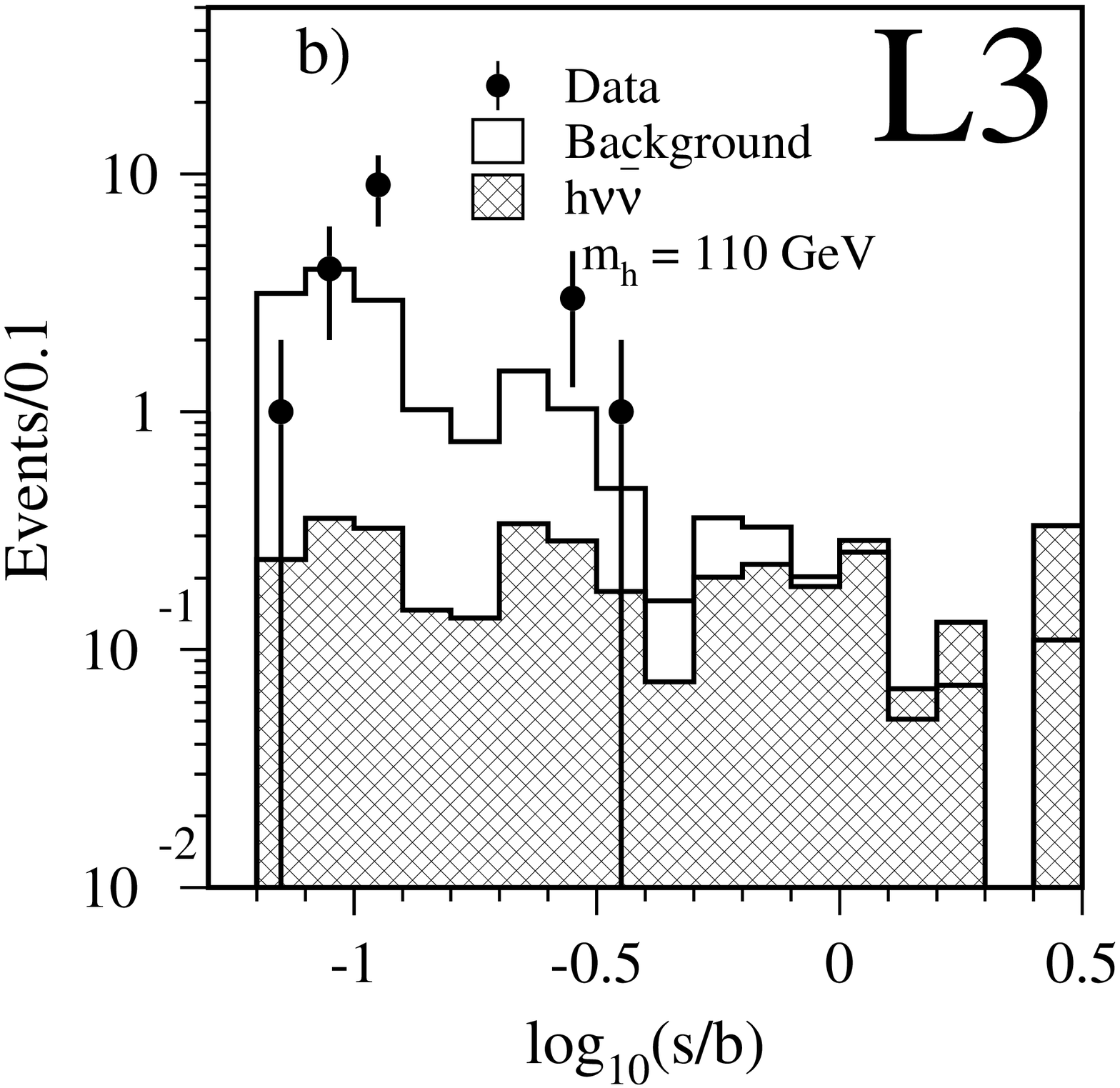} \\
\hspace{-7mm}
\includegraphics*[width=0.47\textwidth]{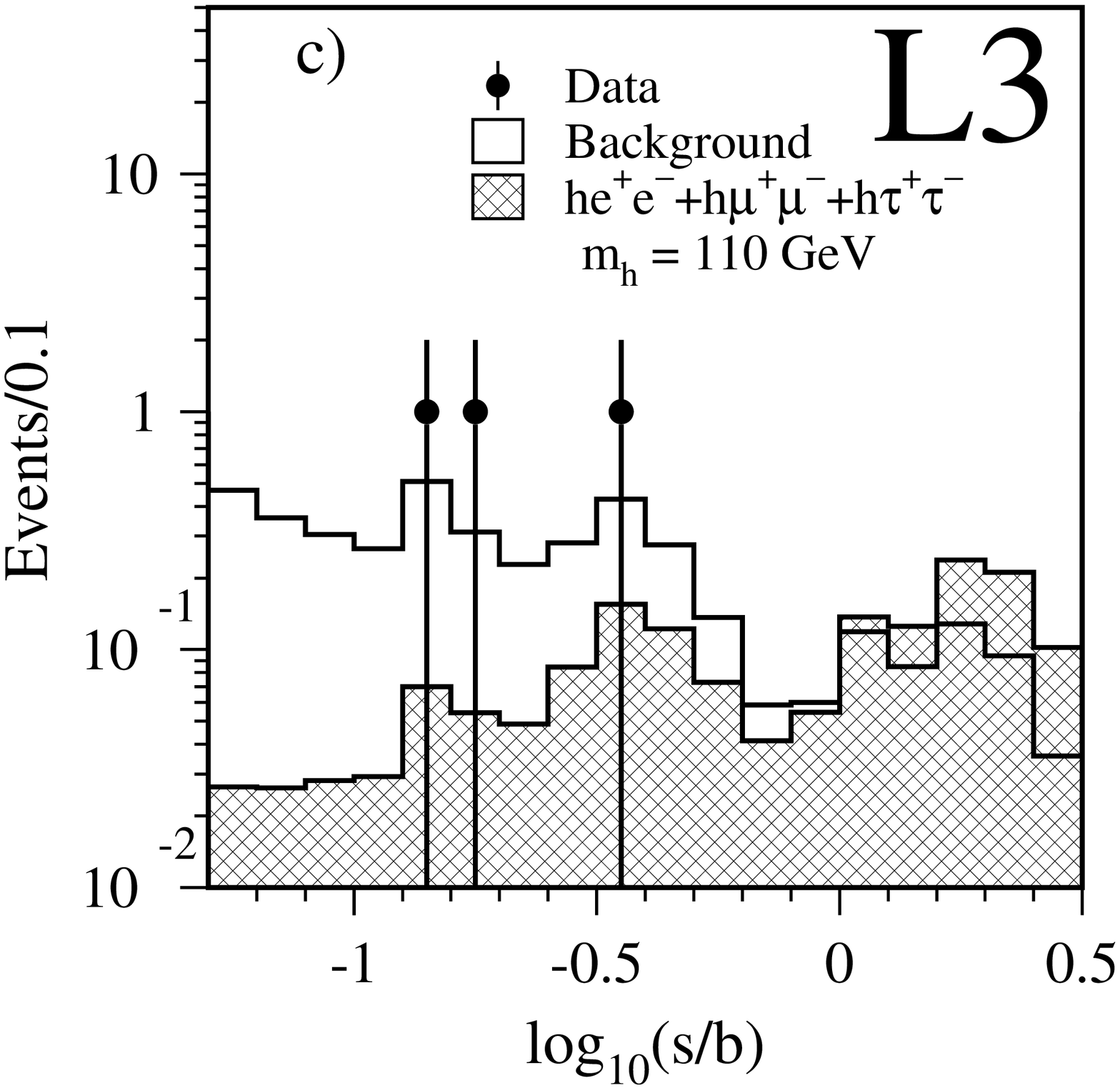} &\\
\end{tabular}
        \caption[]{\label{fig:discriminants}
	  Distributions of the signal over background ratio
	  for events selected in the $\rm e^+e^-\rightarrow hZ$ search by the a)
	  four-jet, b) two jets and missing energy and c) two jets and
	  two lepton analyses. The points indicate data collected at $\sqrt{s}$ $>$ 203\GeV,
          the open histograms represent the expected background
          and the hatched histograms stand for a $\mh$ = 110\GeV signal 
          expected for $\mathrm{\xi^2\times B(h\ra hadrons)}$ =
	  1. Only events with $s/b>0.05$ are shown.}
\end{center}
\end{figure}

For each of the three possible jet pairings, the quantity $\chi^2_{\rm
hZ}=\big( \Sigma - (m_{\h} + m_{\rm Z}) \big)^2/\sigma_{\Sigma}^2 +
\big( \Delta - |m_{\h} - m_{\rm Z}| \big)^2/\sigma_{\Delta}^2 $ is
calculated~\cite{l3_sm_paper}, where $\Sigma$ and $\Delta$ are the
di-jet mass sum and difference, while $\sigma_{\Sigma}$ and
$\sigma_{\Delta}$ are the corresponding resolutions. The pairing which
minimises $ \chi^2_{\rm hZ}$ is chosen and the corresponding value is
used as the final discriminant
variable. Figure~\ref{fig:discriminants}a presents the distributions
of the signal over background ratio in the $\chi^2_{\rm hZ}$ variable
for selected data and Monte Carlo events.  Table~\ref{tab:hz} lists
the numbers of selected and expected events for different $\mh$
hypotheses.

%%%%%%%%%%%%%%%%%%%%%%%%%%%%%%%%%%%%%%%%%%%%%%%%%%%%%%%%%%%%%%
\subsection{Two jets and missing energy}
%%%%%%%%%%%%%%%%%%%%%%%%%%%%%%%%%%%%%%%%%%%%%%%%%%%%%%%%%%%%%%

The signature for h decays into hadrons and Z decays into neutrinos is
a pair of high multiplicity jets, large missing energy and a missing
mass, $m_\mathrm{mis}$, consistent with $\mZ$.  The dominant
backgrounds are the $\mathrm{e^+e^-\rightarrow q\bar{q}(\gamma)}$
process, W pair-production in which only one W decays into hadrons and
Z pair-production with a Z decaying into hadrons and the other into
neutrinos.

\begin{figure}
\begin{center}
\begin{tabular}{cc}
\hspace{-7mm}
\includegraphics*[width=0.5\textwidth]{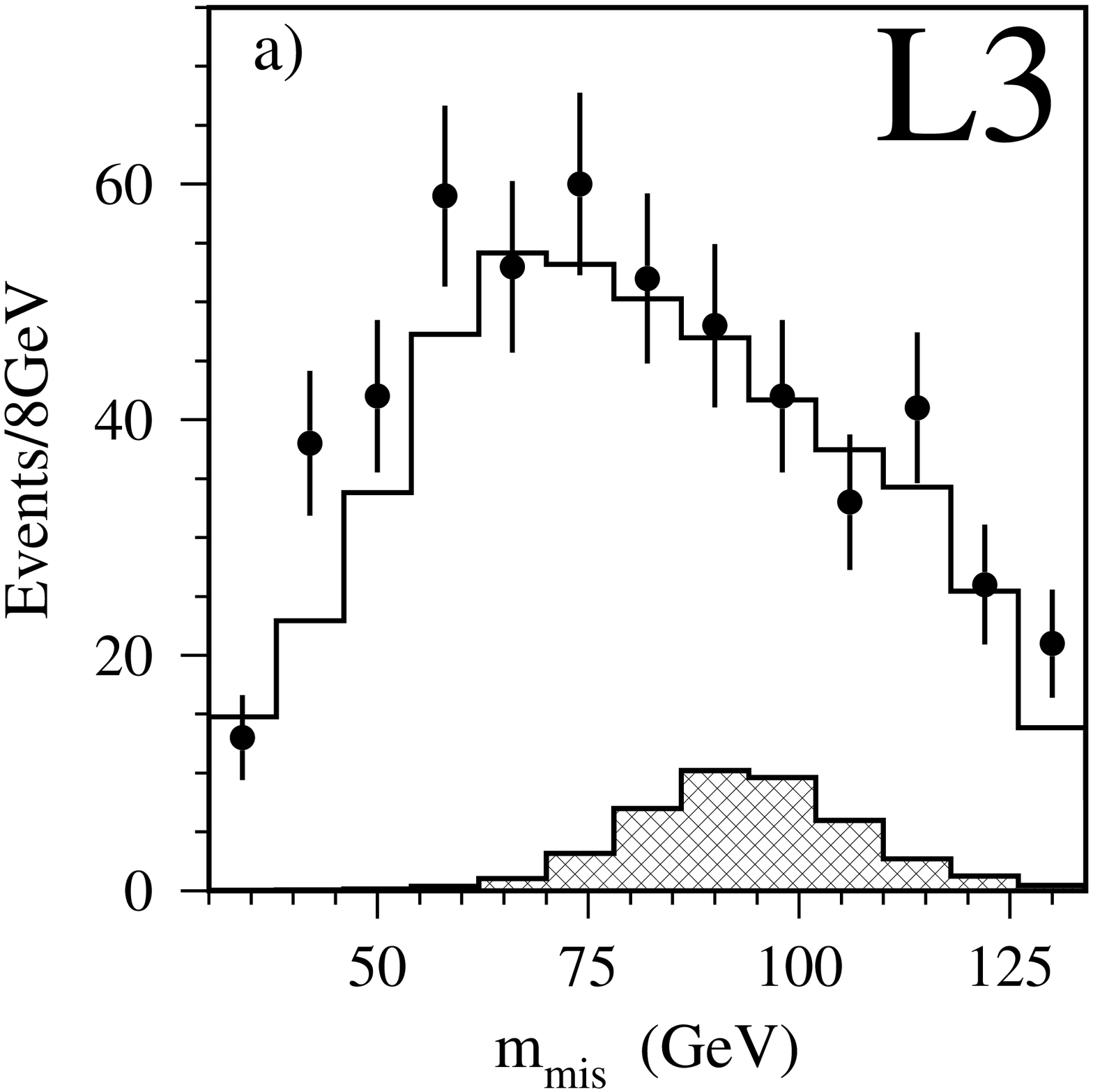} &
\hspace{-0mm}
\includegraphics*[width=0.5\textwidth]{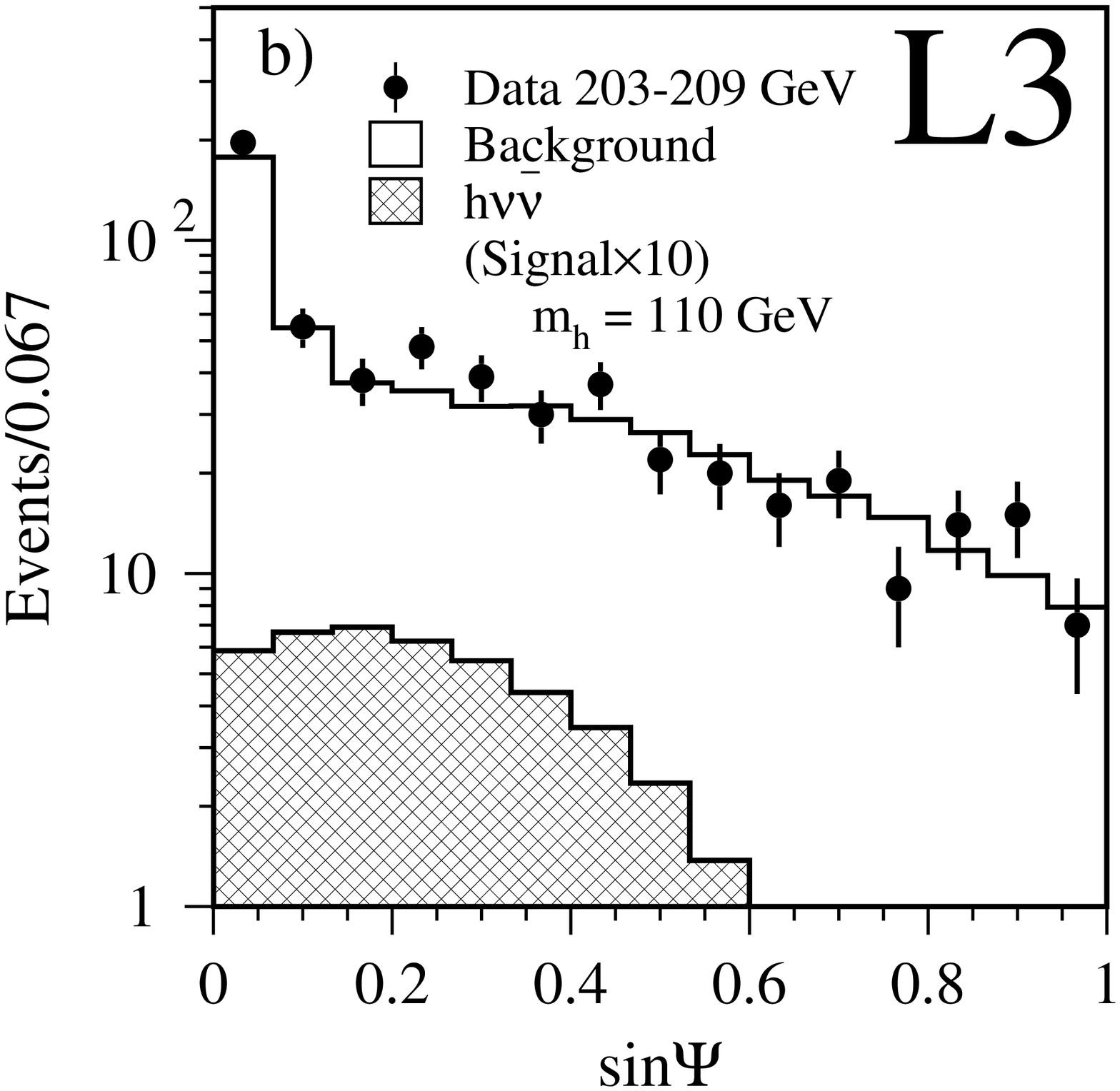} \\
\hspace{-7mm}
\includegraphics*[width=0.5\textwidth]{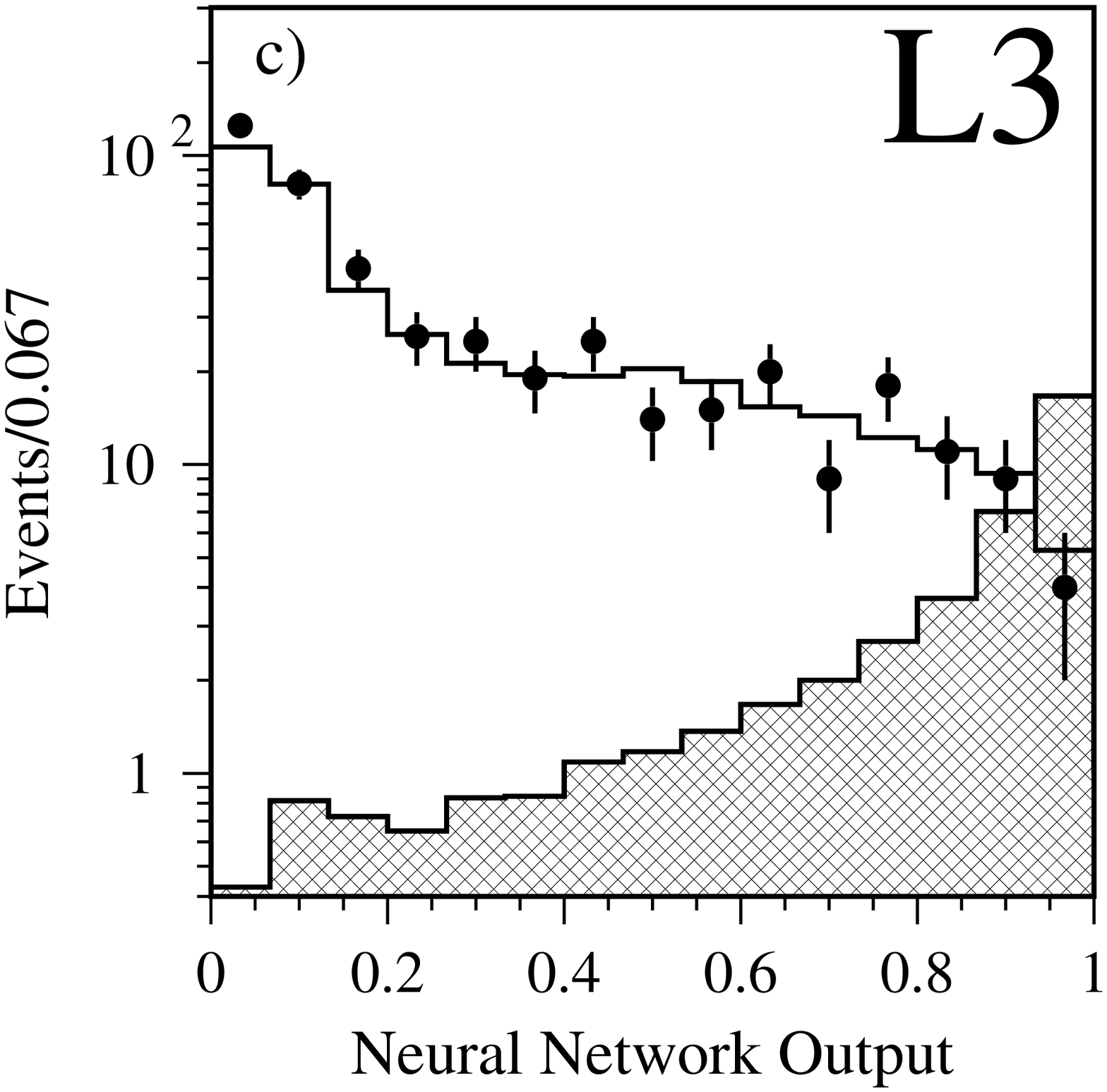} &
\hspace{-0mm}
\includegraphics*[width=0.5\textwidth]{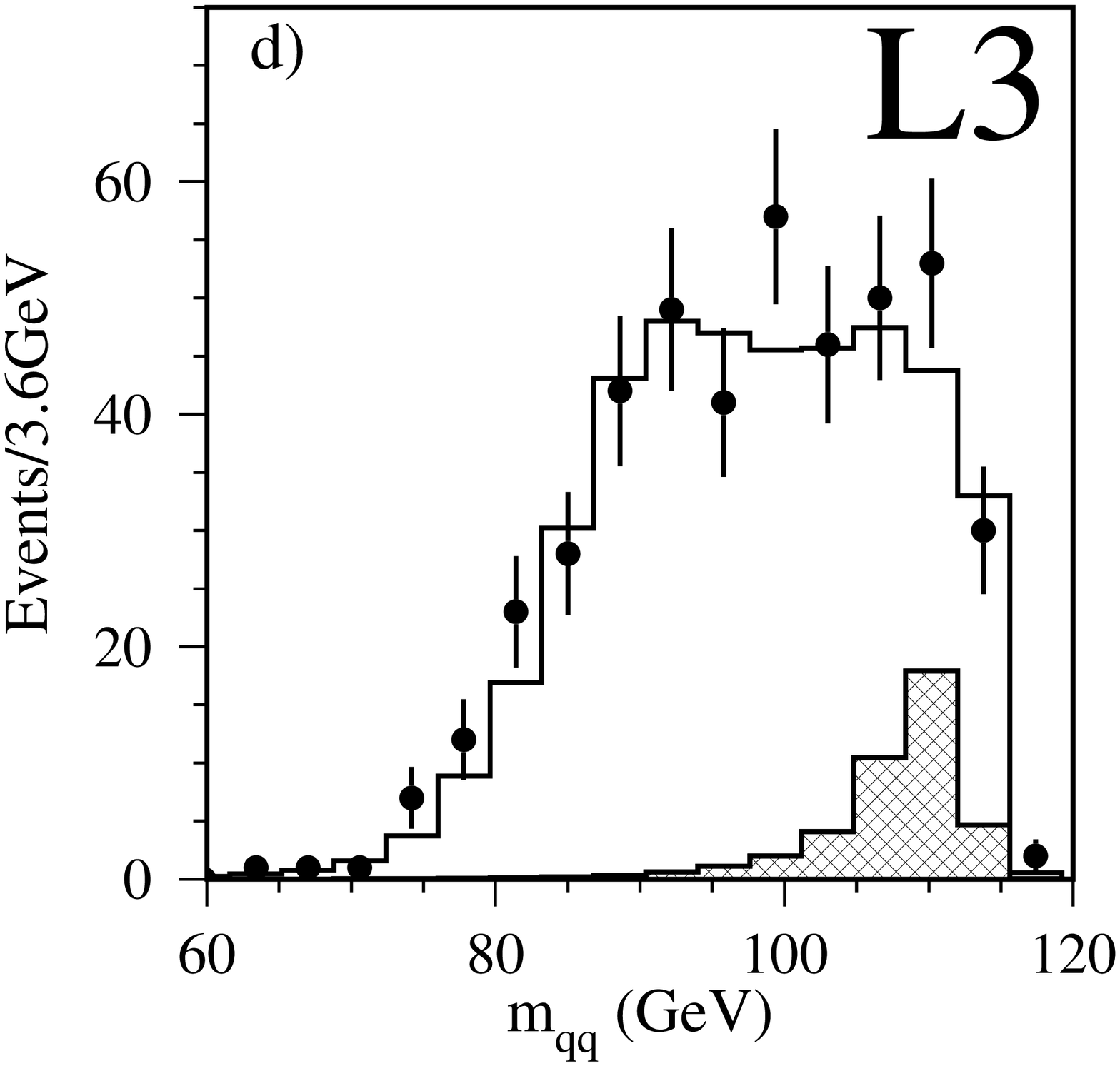} \\
\end{tabular}
        \caption[]{\label{fig:hnn}
        Distributions for the $\rm e^+e^-\rightarrow hZ$ search in the
        two jet and missing energy final state of  
        a) $m_\mathrm{mis}$,
        b) $\sin{\Psi}$,
        c) neural network output and
	d) $m_{\rm qq}$.
        The points indicate data collected at $\sqrt{s}$ $>$ 203\GeV,
        the open histograms represent the expected background
        and the hatched histograms stand for a $\mh$ = 110\GeV signal 
        expected for $\mathrm{\xi^2\times B(h\ra hadrons)}$ = 1, 
        multiplied by a factor of 10.}
\end{center}
\end{figure}

High multiplicity hadronic events are selected with a visible energy,
$E_\mathrm{vis}$, such that $0.25<E_\mathrm{vis}/\sqrt{s}<0.70$. Events
with isolated photons of energy greater than $20\GeV$ are
rejected. The events are forced into two jets using the DURHAM
algorithm and the di-jet mass is required to be greater than 40\GeV to
suppress background from two-photon interactions. Events from the
$\mathrm{e^+e^-\rightarrow q\bar{q}(\gamma)}$ process are suppressed
by requiring $m_\mathrm{mis}>60\GeV$.  In addition, the polar angle,
$\theta$, of the missing momentum must satisfy $|\cos\theta|<0.9$ and
the energy deposited in the very forward calorimeters is required to
be less than 20\GeV. Finally, the sine of the angle $\Psi$ between the
beam axis and the plane spanned by the directions of the two jets must
be greater than 0.025.  Figures~\ref{fig:hnn}a and~\ref{fig:hnn}b
present the distributions of $m_\mathrm{mis}$ and $\sin{\Psi}$ for
data collected at $\sqrt{s}>203\GeV$, expected background and a signal
with $\mh$ = 110\GeV, when all other cuts are applied.

A neural 
network~\cite{jetnet_1} is built from the following variables:
\begin{itemize}
\item $E_\mathrm{vis}$,
\item $m_\mathrm{mis}$,
\item $\sin{\Psi}$,
\item the longitudinal missing momentum,
\item the transverse missing momentum,
\item the absolute value of the cosine of the angle between the two
  jets in the plane transverse to the beam direction,
\item the event thrust,
\item the sum of the jet opening angles after forcing the
event into a three-jet configuration.
\end{itemize}

The distributions of the output of the neural network are presented in 
Figure~\ref{fig:hnn}c. Figure~\ref{fig:hnn}d shows the distributions
of the hadronic mass $m_\mathrm{qq}$,
calculated with a fit which imposes $m_\mathrm{mis}=m_{\rm Z}$. These
two variables are combined into a final discriminant, whose
distributions are presented in Figure~\ref{fig:discriminants}b in
terms of the signal over background ratio. Table~\ref{tab:hz} lists
the  numbers of selected events for different  $\mh$
hypotheses. 

%%%%%%%%%%%%%%%%%%%%%%%%%%%%%%%%%%%%%%%%%%%%%%%%%%%%%%%%%%%%%%
\subsection{Two jets and two leptons}
%%%%%%%%%%%%%%%%%%%%%%%%%%%%%%%%%%%%%%%%%%%%%%%%%%%%%%%%%%%%%%

Different signal topologies correspond to h decays into hadrons and Z
decays into electrons and muons or into tau leptons. For decays into
electrons and muons, the signature is a pair of well isolated
leptons with mass close to $m_{\rm Z}$ and two hadronic jets. In the
case of tau leptons, events with four jets are expected, where two of
the jets are narrow, of low multiplicity, and of unit charge. The
dominant background is due to Z-pair production followed by the
hadronic decay of one Z and the decay into leptons of the other.

The event selection is identical to that used for the same final
states of the Standard Model Higgs search~\cite{l3_sm_paper}. After
this selection, a kinematic fit is applied which imposes four-momentum
conservation and constrains the di-lepton mass to $m_{\rm Z}$. The
mass of the hadronic system after the fit is used as a discriminant to
test different $\mh$ hypotheses. Its distributions in terms of the
signal over background ratio are presented in
Figure~\ref{fig:discriminants}c. The yield of this selection is
presented in Table~\ref{tab:hz}.

%%%%%%%%%%%%%%%%%%%%%%%%%%%%%%%%%%%%%%%%%%%%%%%%%%%%%%%%%%%%%%
\section{Search for  \boldmath$\rm e^+e^-\rightarrow hA$}
%%%%%%%%%%%%%%%%%%%%%%%%%%%%%%%%%%%%%%%%%%%%%%%%%%%%%%%%%%%%%%

The pair-production of \h and \A bosons gives rise to high
multiplicity events with four hadronic jets. The largest backgrounds
are the pair-production of W and Z bosons which decay into hadrons and
the $\mathrm{e^+e^-\rightarrow q\bar{q}(\gamma)}$ process.
High multiplicity events are selected, subjected to a kinematic fit
which enforces four-momentum conservation and forced into four
jets with the DURHAM algorithm. A neural network~\cite{sigmaW} is used
to separate genuine four-jet events from events most likely due to
fermion-pair production.

For each ($\mh$,$\mA$) hypothesis, a likelihood, $L_{\mathrm{hA}}$, is
built~\cite{l3_mssm_paper} to separate the signal from the background
from W- and Z-pair production. It uses the following variables:

\begin{itemize}
\item the maximum energy difference between any two jets, 
\item the minimum jet energy,
\item the probabilities of kinematic fits which impose 
  four-momentum conservation together with the hypotheses of W- or
  Z-pair production,
\item the cosine of the  polar  angle of the di-jet system which
  best fits the hA pair-production hypothesis,
\item the cosine of the polar angle, $\Theta_{\rm W^+}$, at which the positive
  charged\footnote{Charge assignment is based on jet-charge
  techniques~\cite{TGC}.} boson is produced for the di-jet system which
  best fits the W-pair production hypothesis,
\item $y_{34}$,
\item the absolute value of the cosine of the polar angle,
 $\Theta_{\rm T}$, of the thrust axis.
\end{itemize}

\begin{figure}
\begin{center}
\begin{tabular}{cc}
\hspace{-7mm}
\includegraphics*[width=0.47\textwidth]{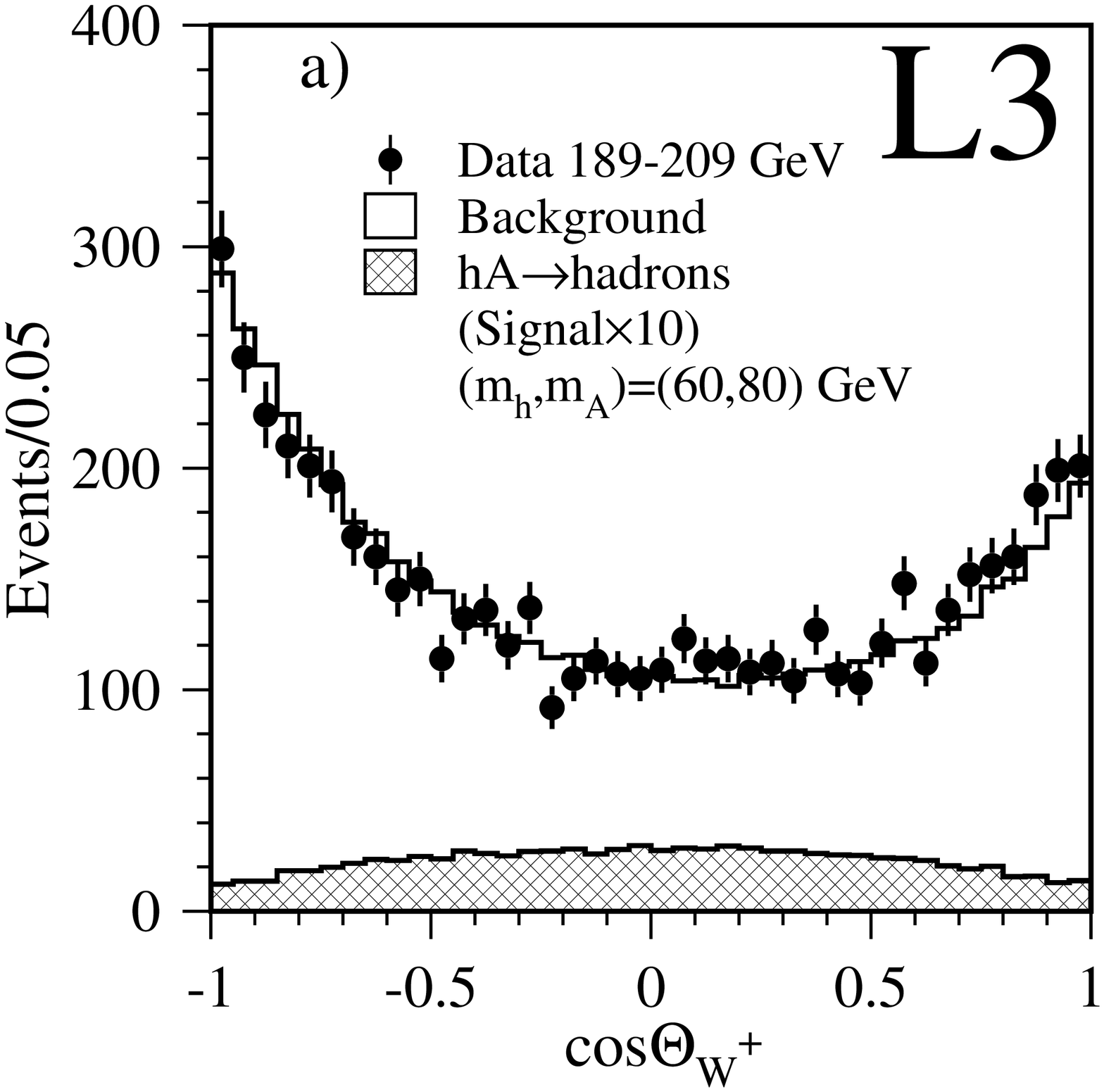} &
\hspace{-0mm}
\includegraphics*[width=0.47\textwidth]{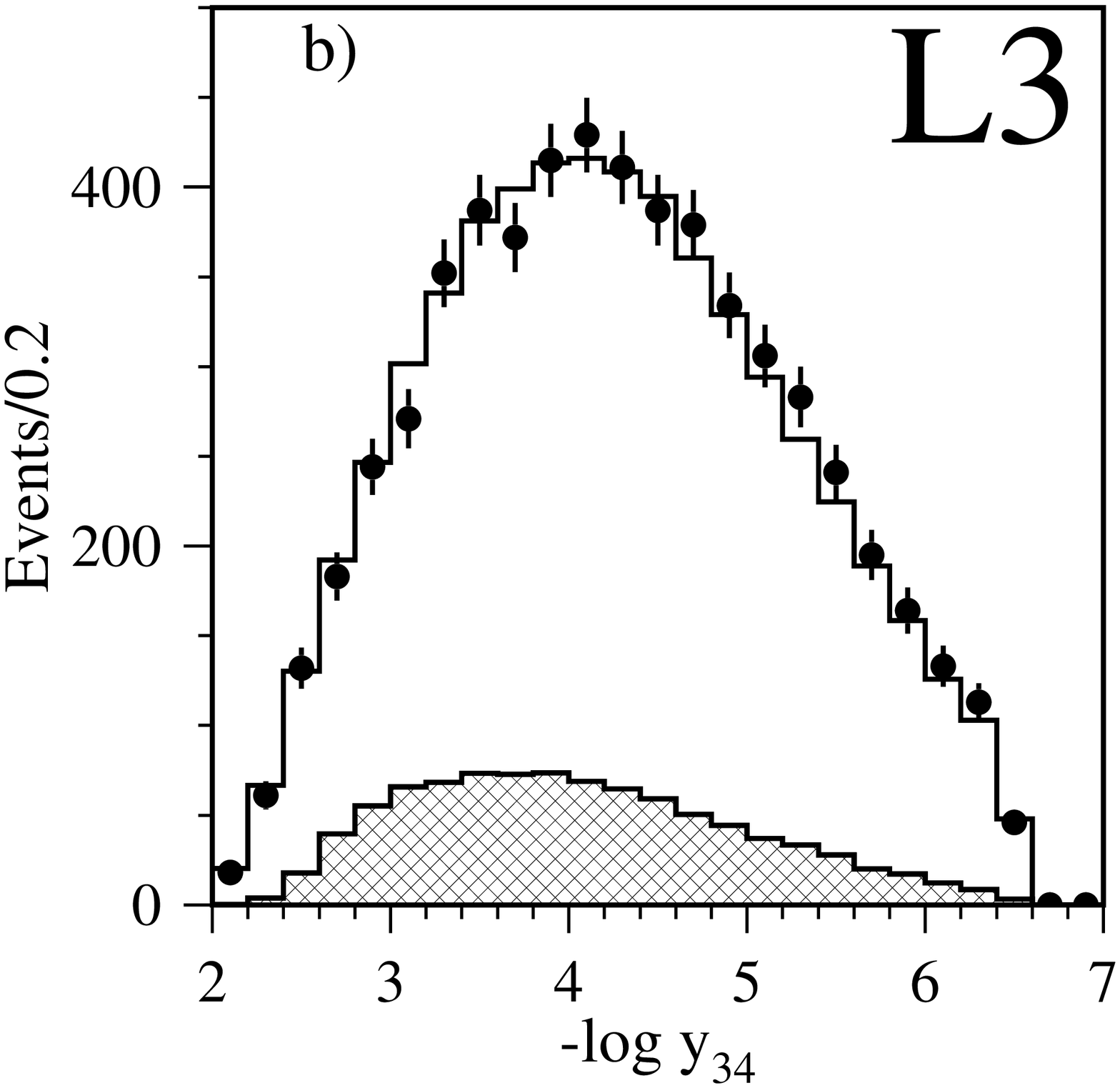} \\
\hspace{-7mm}
\includegraphics*[width=0.47\textwidth]{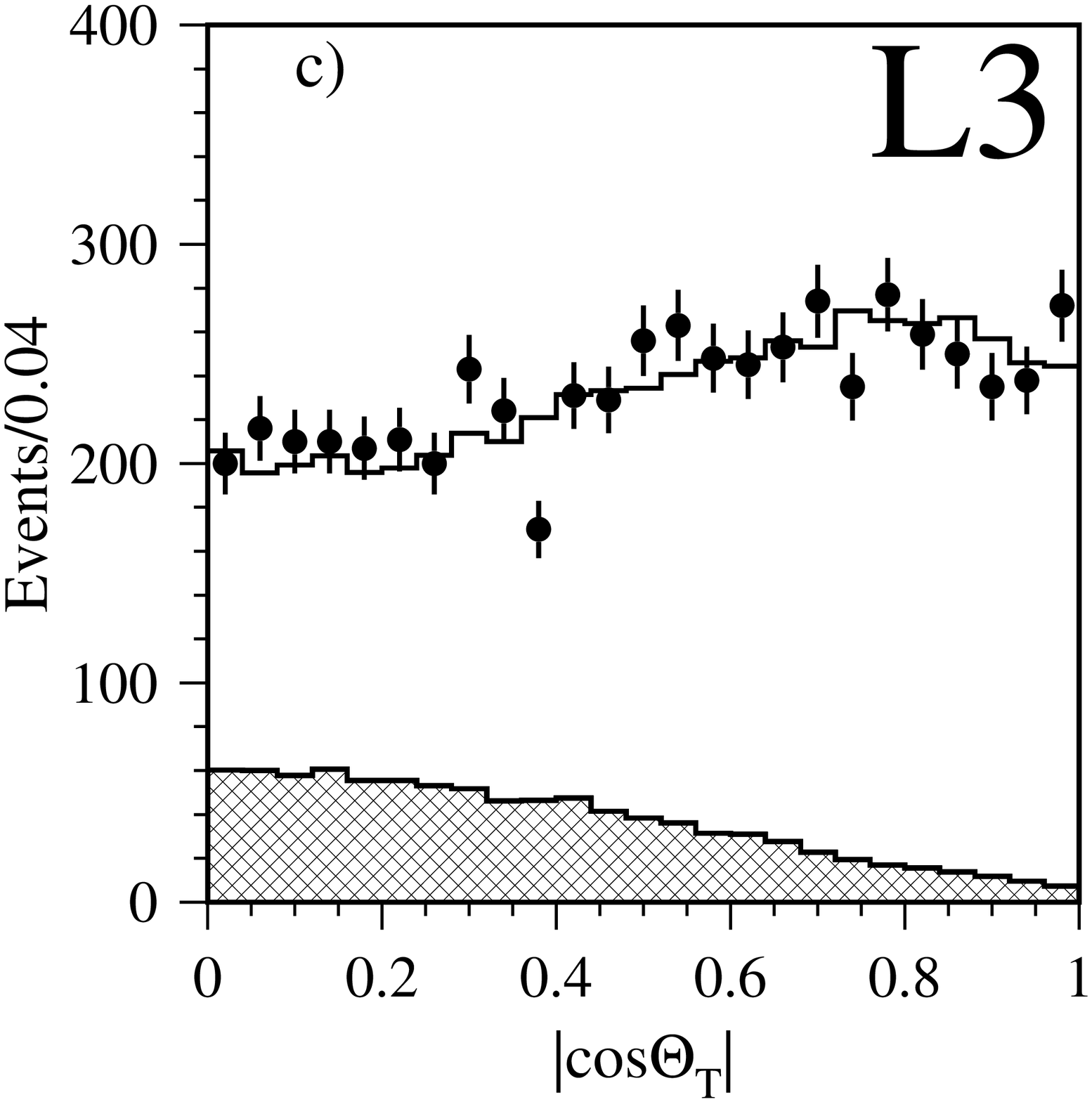} &
\hspace{-0mm}
\includegraphics*[width=0.47\textwidth]{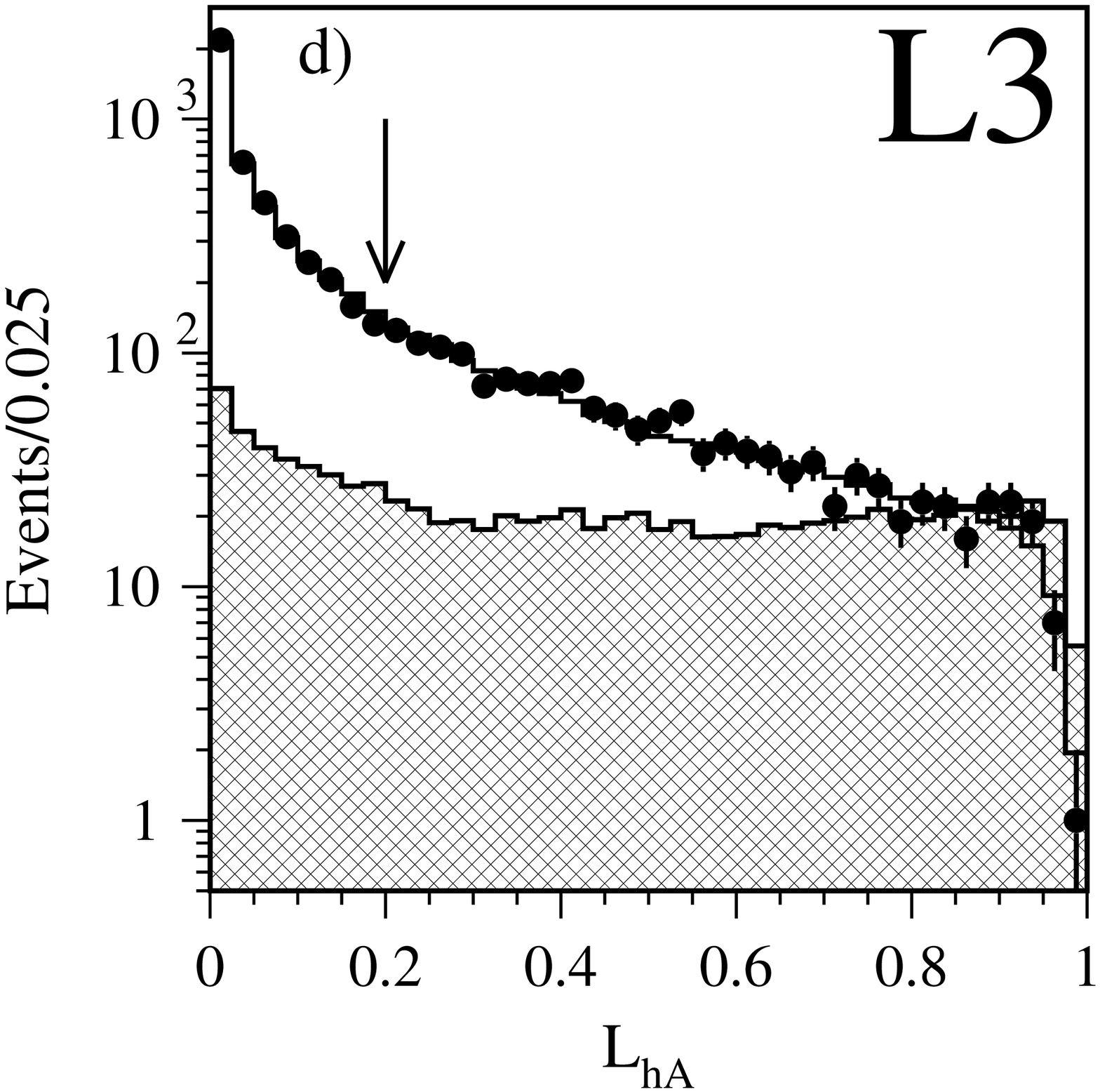} \\
\end{tabular}
        \caption[]{\label{fig:qqqq}
        Distributions for the $\rm e^+e^-\rightarrow hA$ search of:  
	a) $\cos{\Theta_{\rm W^+}}$,
        b) $-\log{y_{34}}$,
	c) $|\cos{\Theta_{\rm T}}|$,
        d) $L_\mathrm{hA}$.
        The points indicate the data,
        the open histograms represent the expected background
        and the hatched histograms stand for a ($\mh$,$\mA$) = (60,80)\GeV signal 
        expected for $\mathrm{\eta^2\times B(hA\ra hadrons)}$ = 1, 
        multiplied by a factor of 10. The arrow in d) indicates
        the position of the cut.}
\end{center}
\end{figure}

\begin{figure}
\begin{center}
\begin{tabular}{cc}
\hspace{-7mm}
\includegraphics*[width=0.4\textwidth]{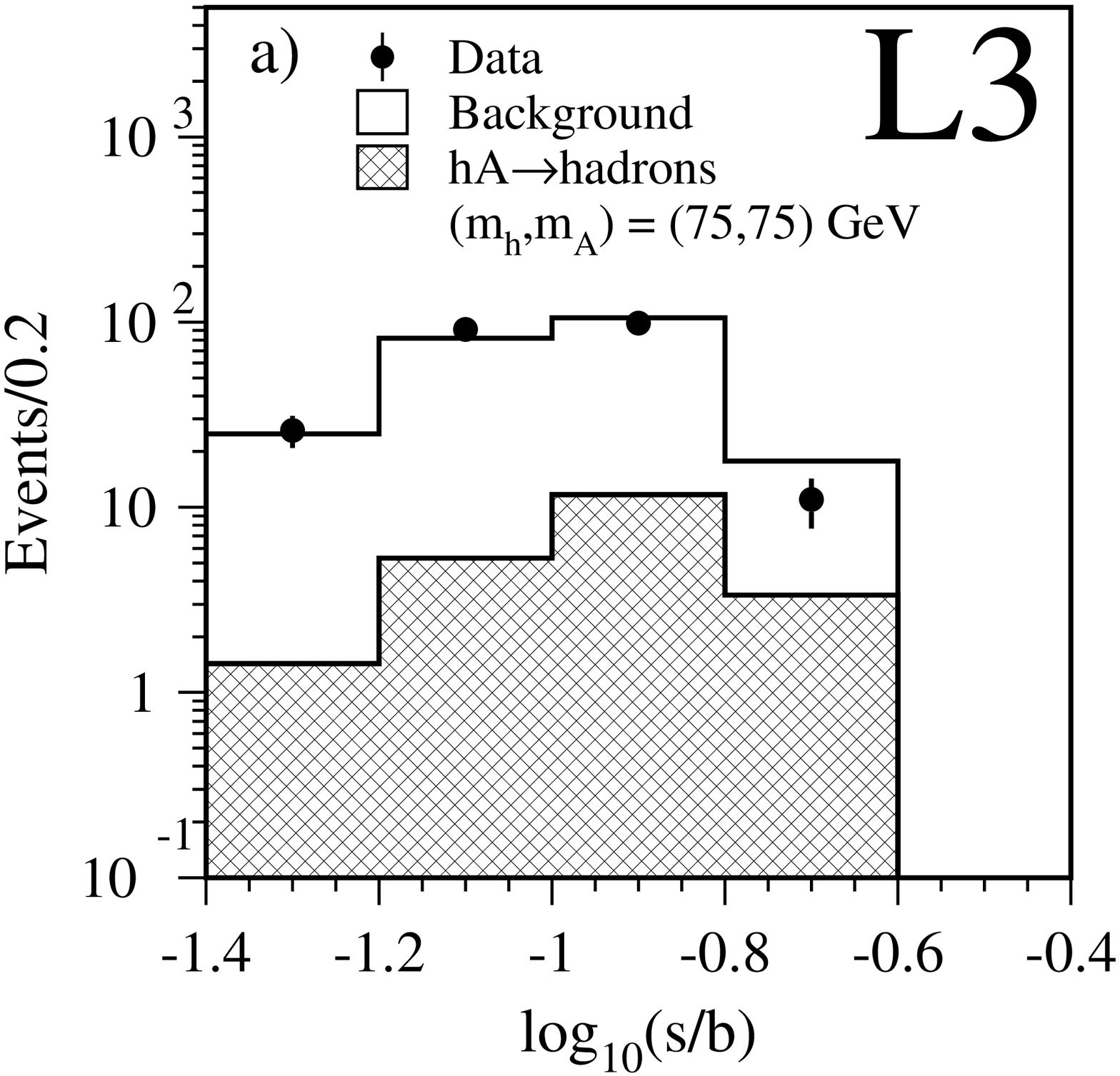} &
\hspace{-0mm}
\includegraphics*[width=0.4\textwidth]{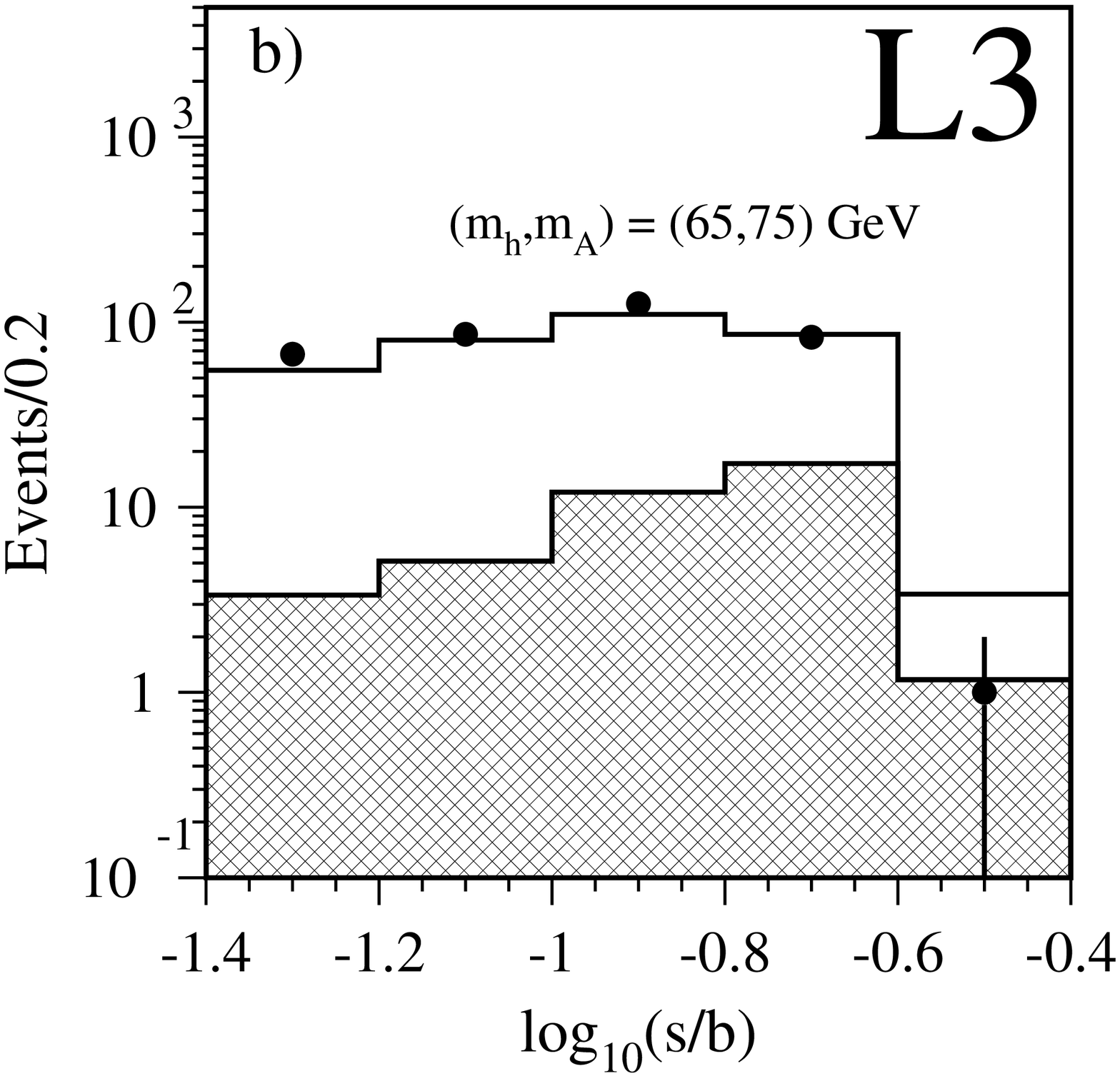} \\
\hspace{-7mm}
\includegraphics*[width=0.4\textwidth]{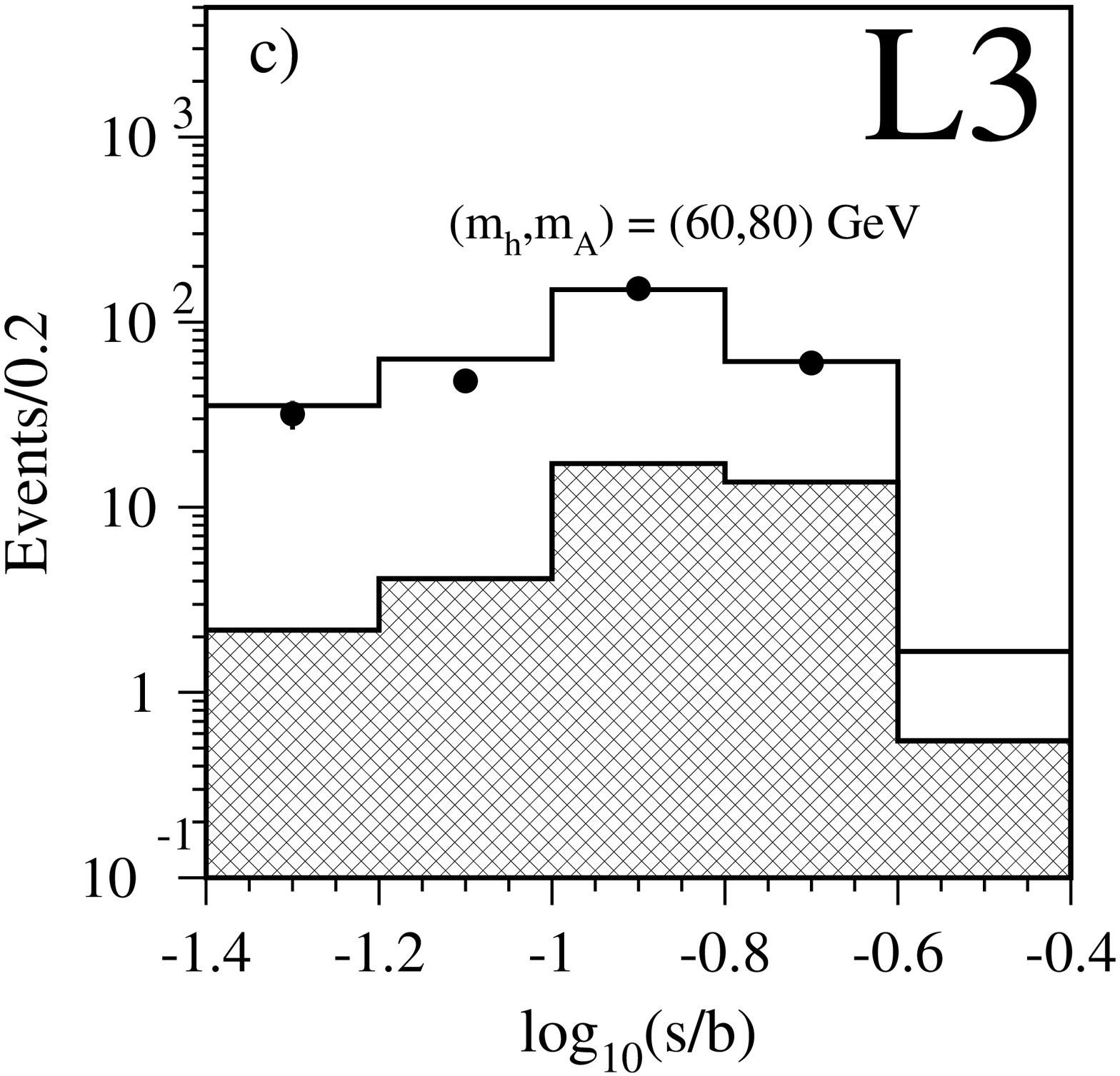} &
\hspace{-0mm}
\includegraphics*[width=0.4\textwidth]{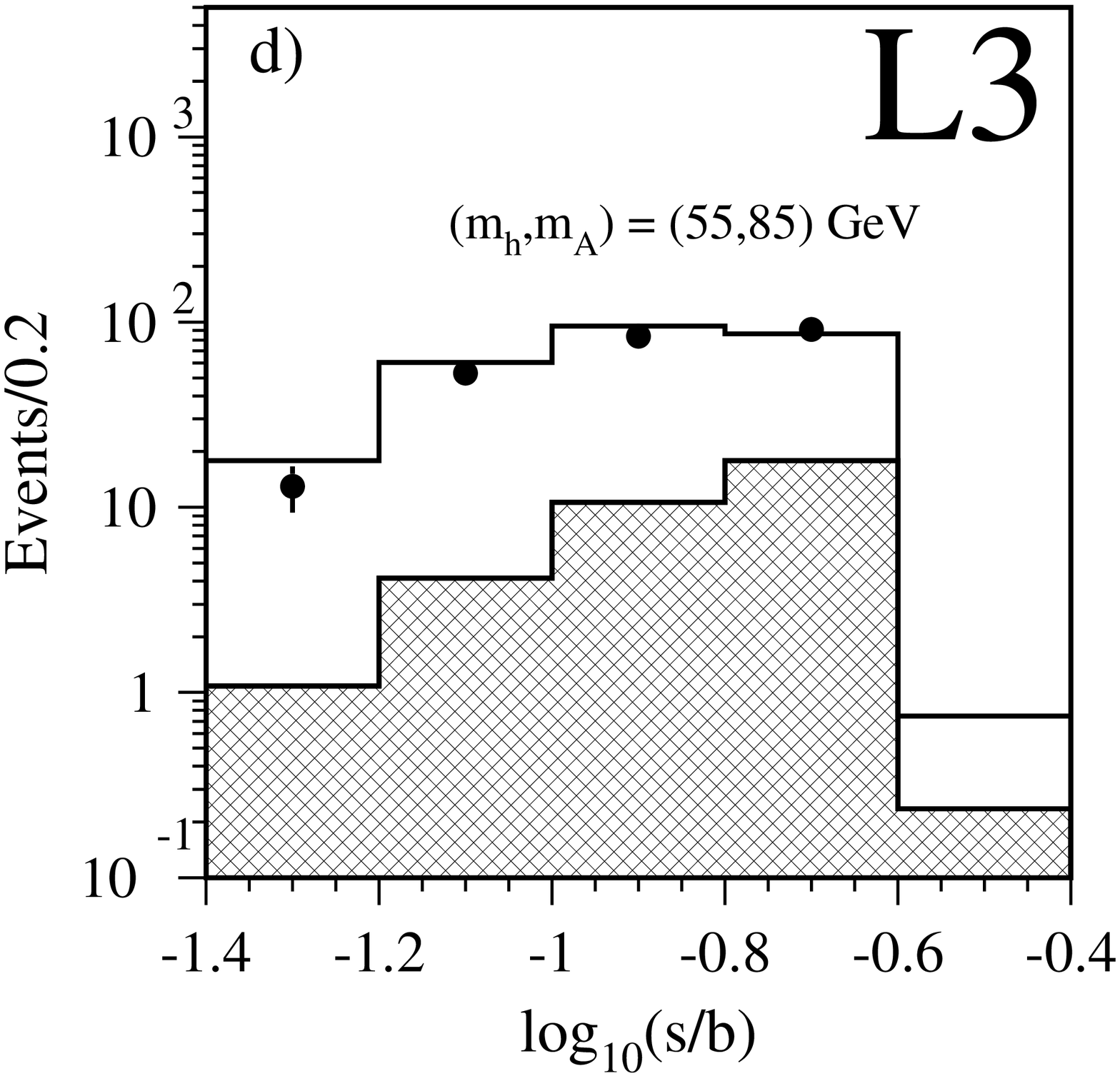} \\
\hspace{-7mm}
\includegraphics*[width=0.4\textwidth]{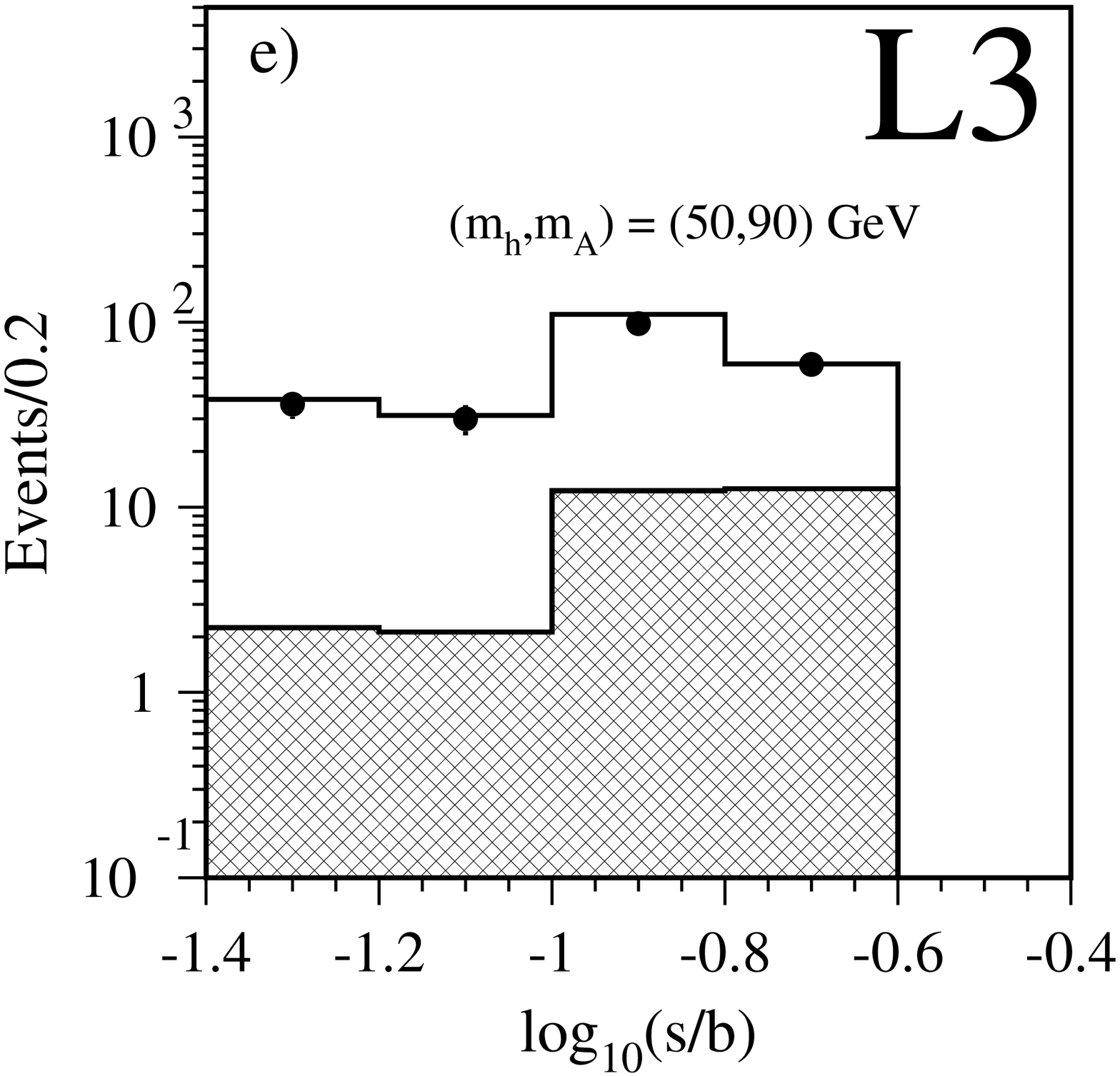} &
\hspace{-0mm}
\includegraphics*[width=0.4\textwidth]{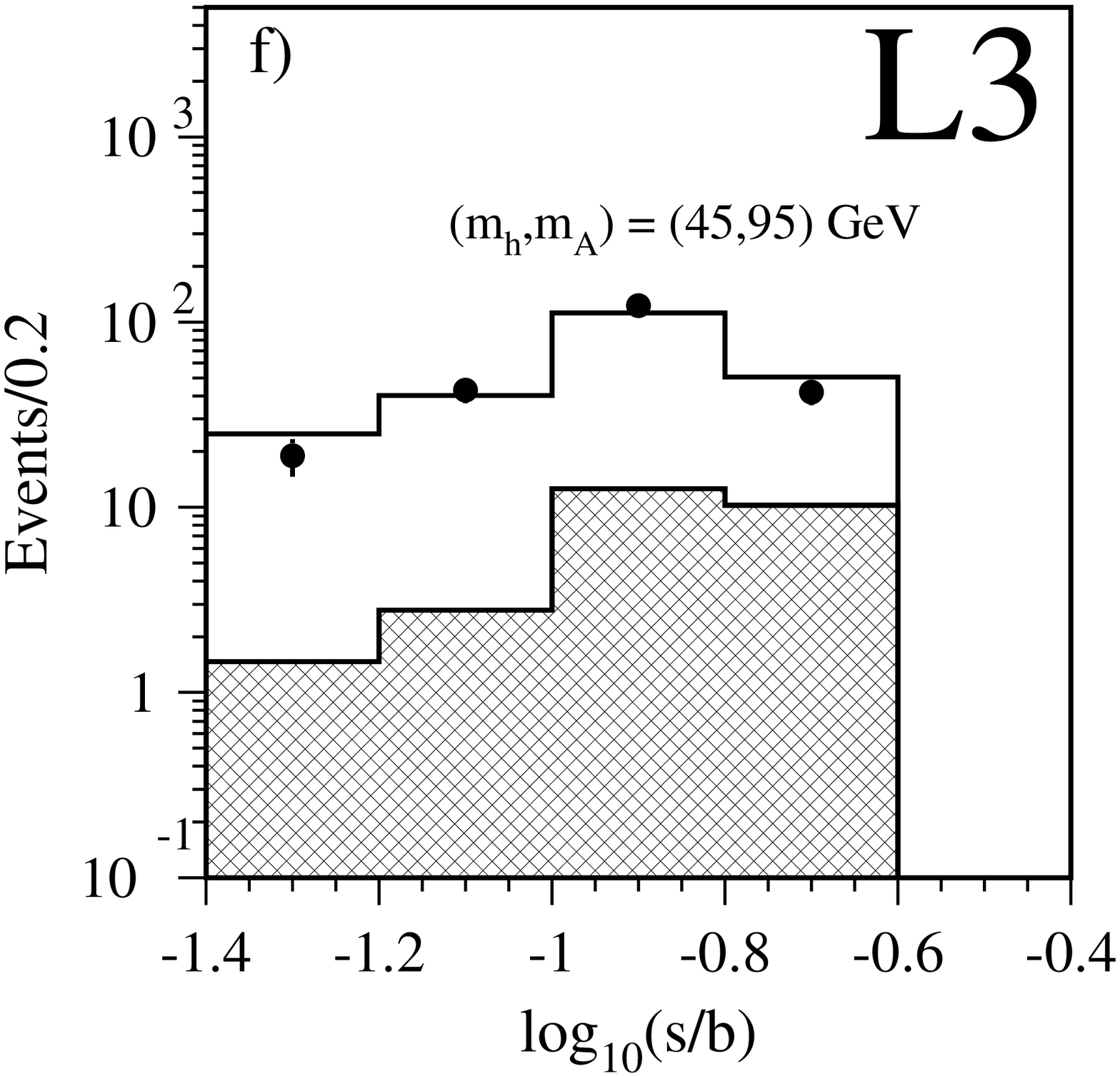} \\
\end{tabular}
        \caption[]{\label{fig:hA-disc}
	  Distributions of the signal over background ratio
	  for events selected by the  $\rm e^+e^-\rightarrow hA$
	  search for different  ($\mh$,$\mA$) mass hypotheses: 
	  a) (75,75)\GeV,
	  b) (65,75)\GeV,
	  c) (60,80)\GeV,
	  d) (55,85)\GeV,
	  e) (50,90)\GeV and
	  f) (45,95)\GeV.
        The points indicate the data,
        the open histograms represent the expected background
        and the hatched histograms stand for a signal of the given
	  ($\mh$,$\mA$) hypothesis
        expected for $\mathrm{\eta^2\times B(hA\ra hadrons)}$ = 1.}
\end{center}
\end{figure}

Figure~\ref{fig:qqqq} shows the distributions of the last three
variables and of $L_\mathrm{hA}$ for data, the expected background and
the signal corresponding to the Higgs boson mass hypothesis
($\mh$,$\mA$) = (60,80)\GeV. A cut on $L_\mathrm{hA}$ is applied, which
depends on $\sqrt{s}$ and on the ($\mh$,$\mA$) hypothesis, typically
around 0.2.  The remaining events are tested for consistency with a
given ($\mh$,$\mA$) hypothesis by means of the variable $ \chi^2_{\rm
hA}$~\cite{l3_mssm_paper}, defined analogously to $ \chi^2_{\rm
hZ}$. The pairing which minimises the value of $\chi^2_{\rm hA}$ is
chosen.  For each event and each ($\mh$,$\mA$) hypothesis, the value
of the signal over background ratio of the variable
$\chi^2_\mathrm{hA}$ is calculated.  The distributions of these ratios
are presented in Figure~\ref{fig:hA-disc} for different mass
hypotheses.

Table~\ref{tab:ha} reports the numbers of observed events, expected
background and expected signal events for several Higgs boson mass
hypotheses, together with selection efficiencies.

%%%%%%%%%%%%%%%%%%%%%%%%%%%%%%%%%%%%%%%%%%
\section{Results}
%%%%%%%%%%%%%%%%%%%%%%%%%%%%%%%%%%%%%%%%%%

Table~\ref{tab:hz} shows the result of the combination of the
different channels of the $\rm e^+e^- \rightarrow hZ$ search. The
observed number of events agrees with the Standard Model
expectations. No significant excess is observed either in the $\rm
e^+e^- \rightarrow hZ$ search or in the $\rm e^+e^- \rightarrow hA$
search, which is summarised in Table~\ref{tab:ha}. Limits on the $\xi$ and
$\eta$ couplings are extracted as a function of $m_{\rm h}$ and
$m_{\rm A}$ from the distributions of the signal over background
ratios derived from the final discriminant variables. The
log-likelihood ratio technique~\cite{lnq} is used for the combination
of the different channels of the $\rm e^+e^- \rightarrow hZ$ search
and to derive all the limits.  For each final state, among the three
possible decays of the h and A bosons into $\mathrm{b\bar{b}}$,
$\mathrm{c\bar{c}}$ and $\mathrm{gg}$, the case with the lowest
efficiency is considered.

Several sources of systematic uncertainties are investigated and their
impact on the signal efficiency and the determination of the
background level is assesed. The limited Monte Carlo statistics
affects the signal by around 2\% and the background by around 5\%,
depending on the final state. The selection criteria are varied
within the resolution of the corresponding variables yielding an
uncertainty from the selection procedure around 2\% on the signal and
from 3\% to 6\% on the background.  Lepton identification criteria
contribute to this source with an additional 1\% for the signal and
2\% for the background. The expected background level has an
uncertainty up to 5\%, depending on the final state, due to the
uncertainty in the calculation of the cross sections of background
processes. 

Particular care is payed to validate the accuracy of the simulation of
gluon jets. A reference sample of three-jet events, from the
$\mathrm{e^+e^-\ra q\bar{q}g(\gamma)}$ process, is selected and the
jet with the smallest energy in the rest frame of the hadronic system
is taken as the gluon jet.  The distributions of the most important
gluon jet characteristics such as jet broadening, boosted sphericity
and charged track multiplicity are compared for data and Monte Carlo
samples. The latter, for instance, is found to be on average
overestimated by the simulations and is not considered as input to the
likelihoods and the neural networks. From this comparison, an
additional systematic uncertainty is assigned as 1.5\% for the signal
and 2\% for the background.

The overall systematic uncertainties depend on the search channel and
are estimated to range between 2\% and 4\% for the signal efficiencies
and between 4\% and 8\% for the background levels. They are included
in the derivation of the limits. For $\xi^2=1$ they lower the
sensitivity to $\mh$ by about 0.8\GeV and for $\eta^2=1$ and $\mh=\mA$
by about 0.7\GeV.

\begin{figure}
\begin{center}
\includegraphics*[width=1.0\textwidth]{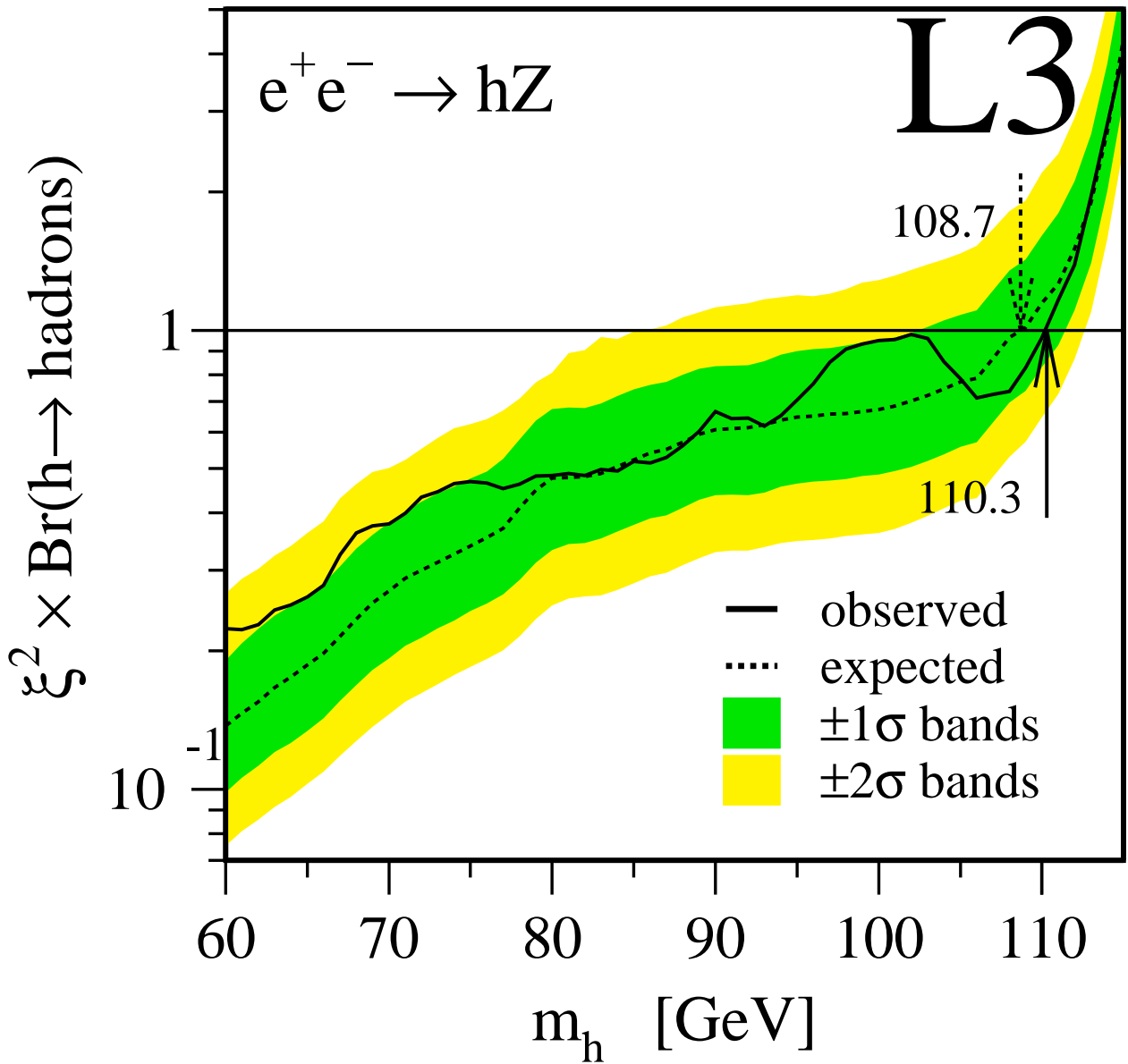}
        \caption[]{\label{fig:hZ-xsec}
        The 95\% confidence level upper limit on 
        $\mathrm{\xi^2\times Br(h\ra hadrons)}$ as a function 
        of  $m_{\rm h}$. The solid line indicates the
        observed limit and the dashed line stands for the 
        median expected limit. The shaded areas show the $1\sigma$
        and $2\sigma$ intervals centered on the median expected
        limit. The observed and expected limits on $m_{\rm h}$ for
        $\mathrm{\xi^2\times Br(h\ra hadrons)}=1$ are also shown.
        }
\end{center}
\end{figure}

Figure~\ref{fig:hZ-xsec} shows the 95\% confidence level (CL) upper
limit on $\mathrm{\xi^2\times B(h\ra hadrons)}$ as a function of
$m_{\rm h}$. The expected limit and the 68.3\% and 95.4\% probability
bands expected in the absence of a signal are also displayed and
denoted as $1\sigma$ and $2\sigma$, respectively.  For
$\mathrm{\xi^2\times B(h\ra hadrons)}= 1$, {\it i.e.}  for a cross
section equivalent to the Standard Model one and a Higgs boson
decaying into hadrons, a 95\% CL lower limit of 110.3$\GeV$ is set on
$\mh$.  The expected limit is 108.7$\GeV$.

\begin{figure}
\begin{center}
\begin{tabular}{cc}
\hspace{-7mm}
\includegraphics*[width=0.4\textwidth]{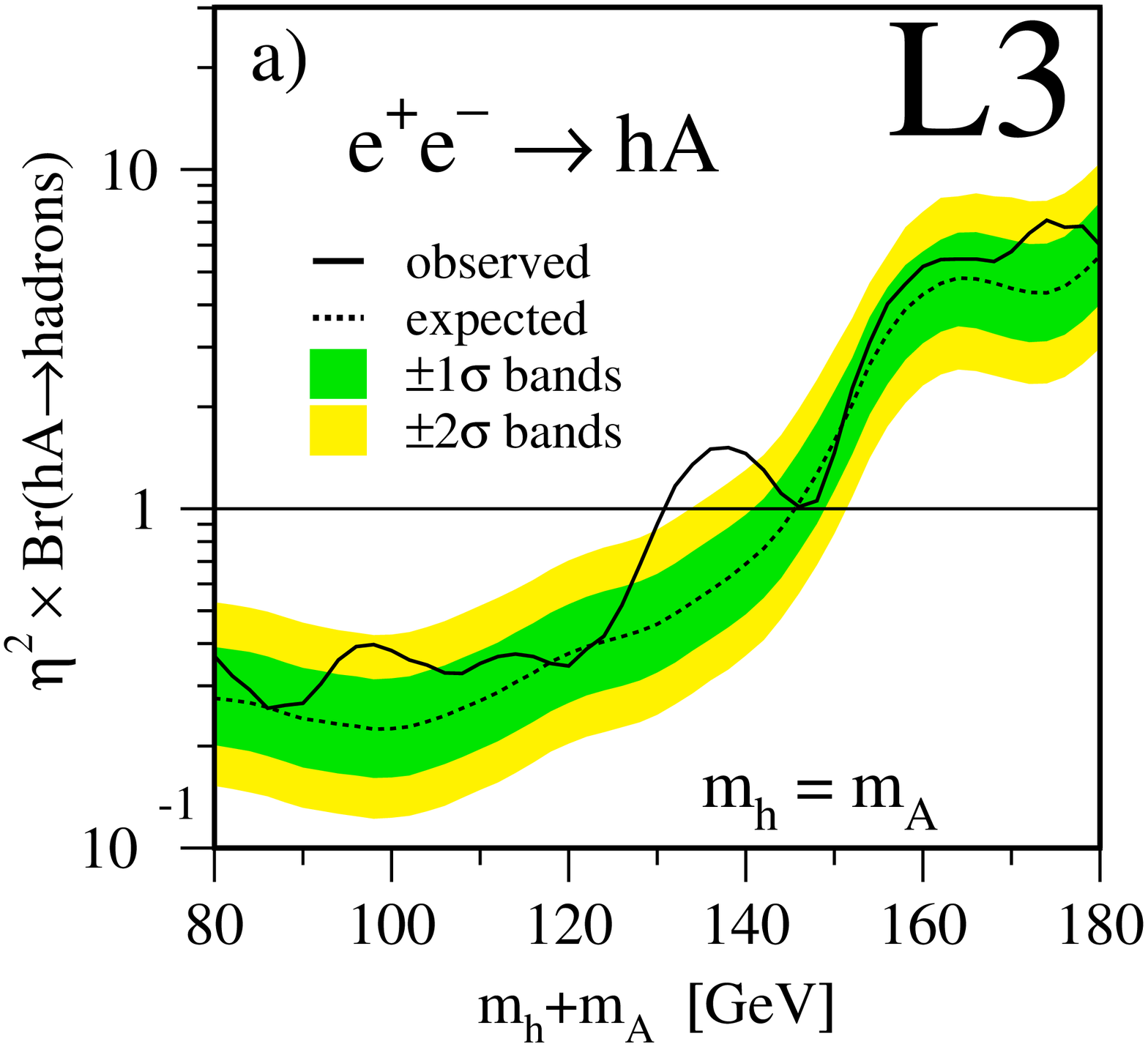} &
\hspace{-0mm}
\includegraphics*[width=0.4\textwidth]{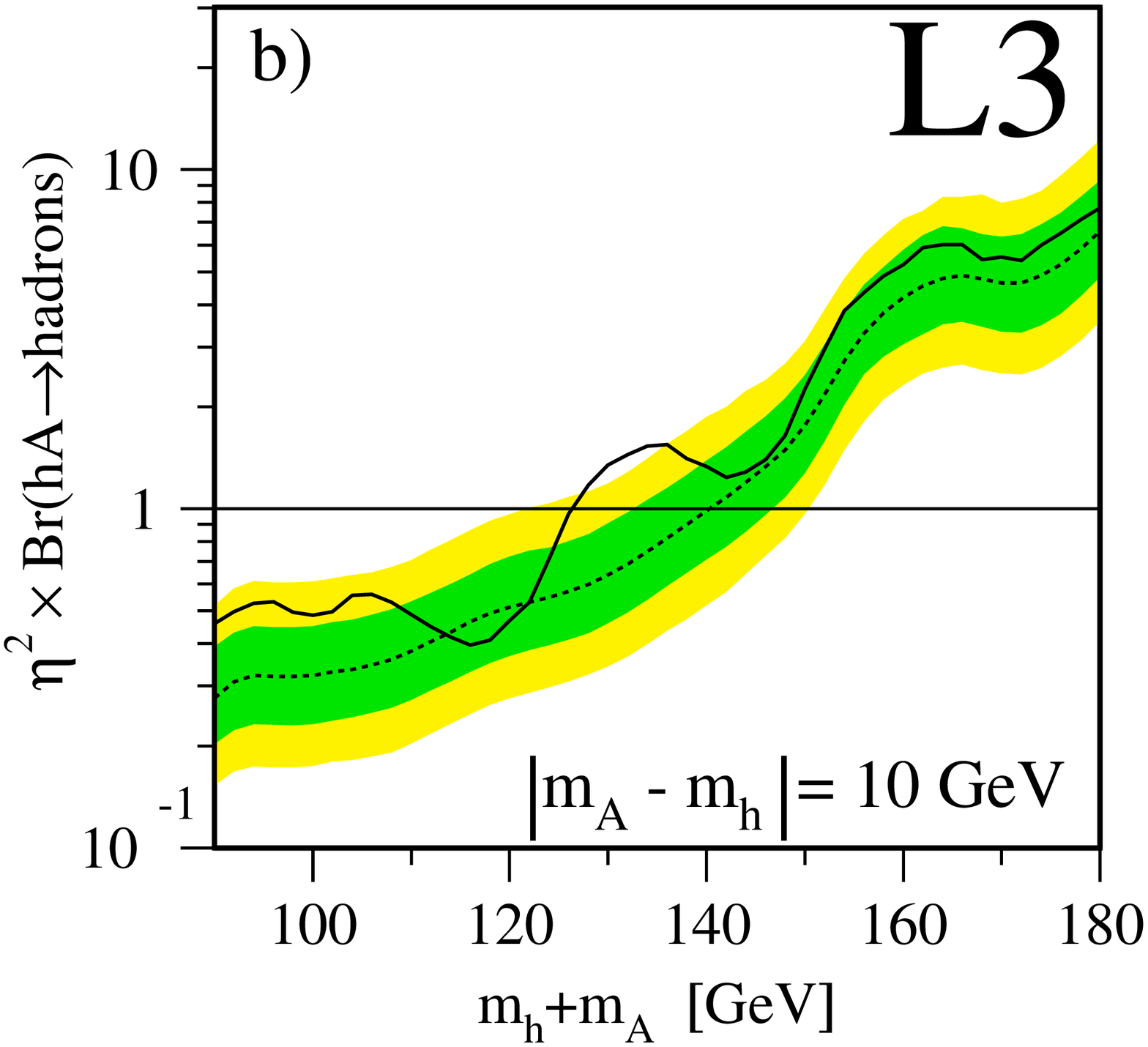} \\
\hspace{-7mm}
\includegraphics*[width=0.4\textwidth]{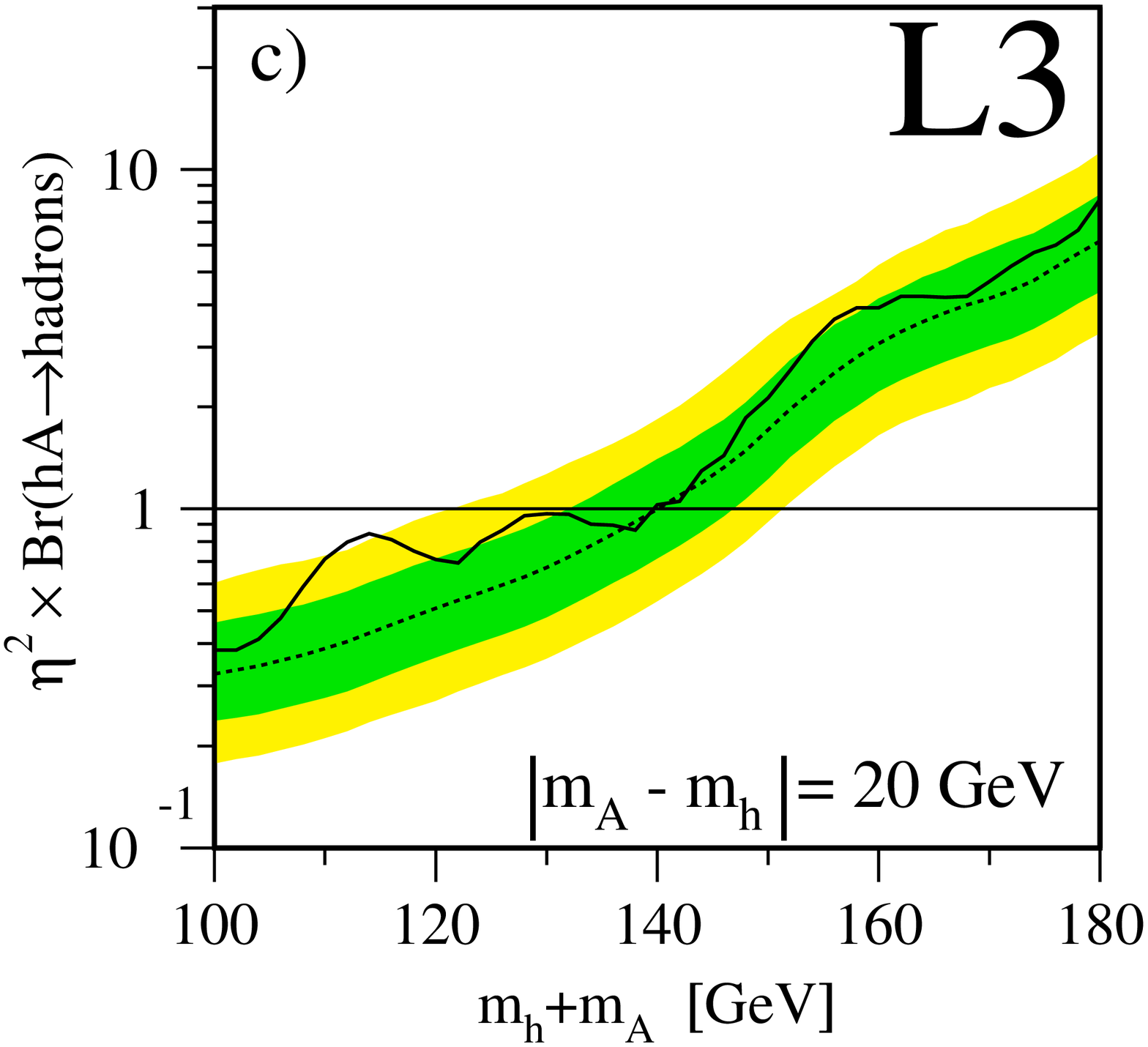} &
\hspace{-0mm}
\includegraphics*[width=0.4\textwidth]{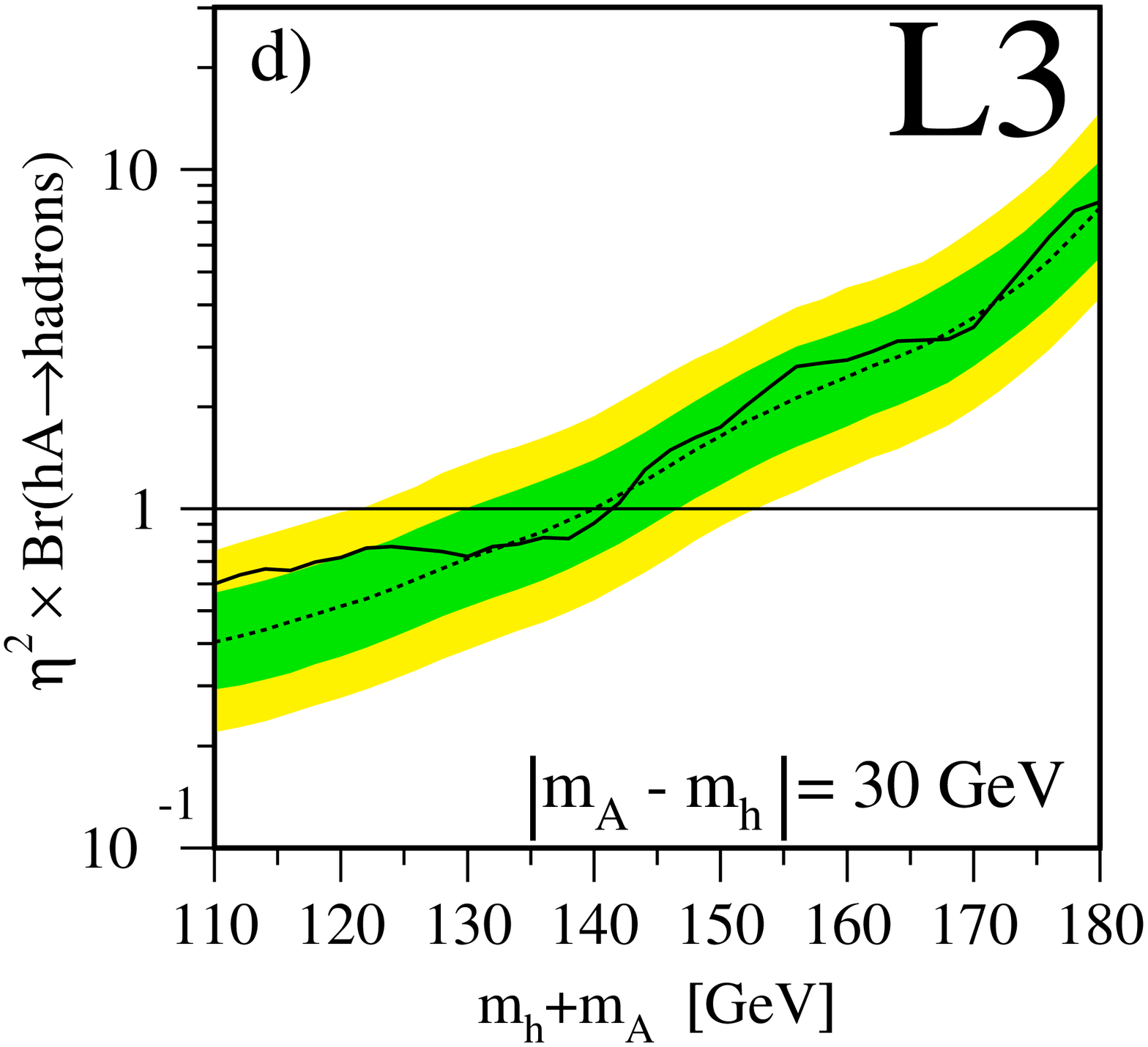} \\
\hspace{-7mm}
\includegraphics*[width=0.4\textwidth]{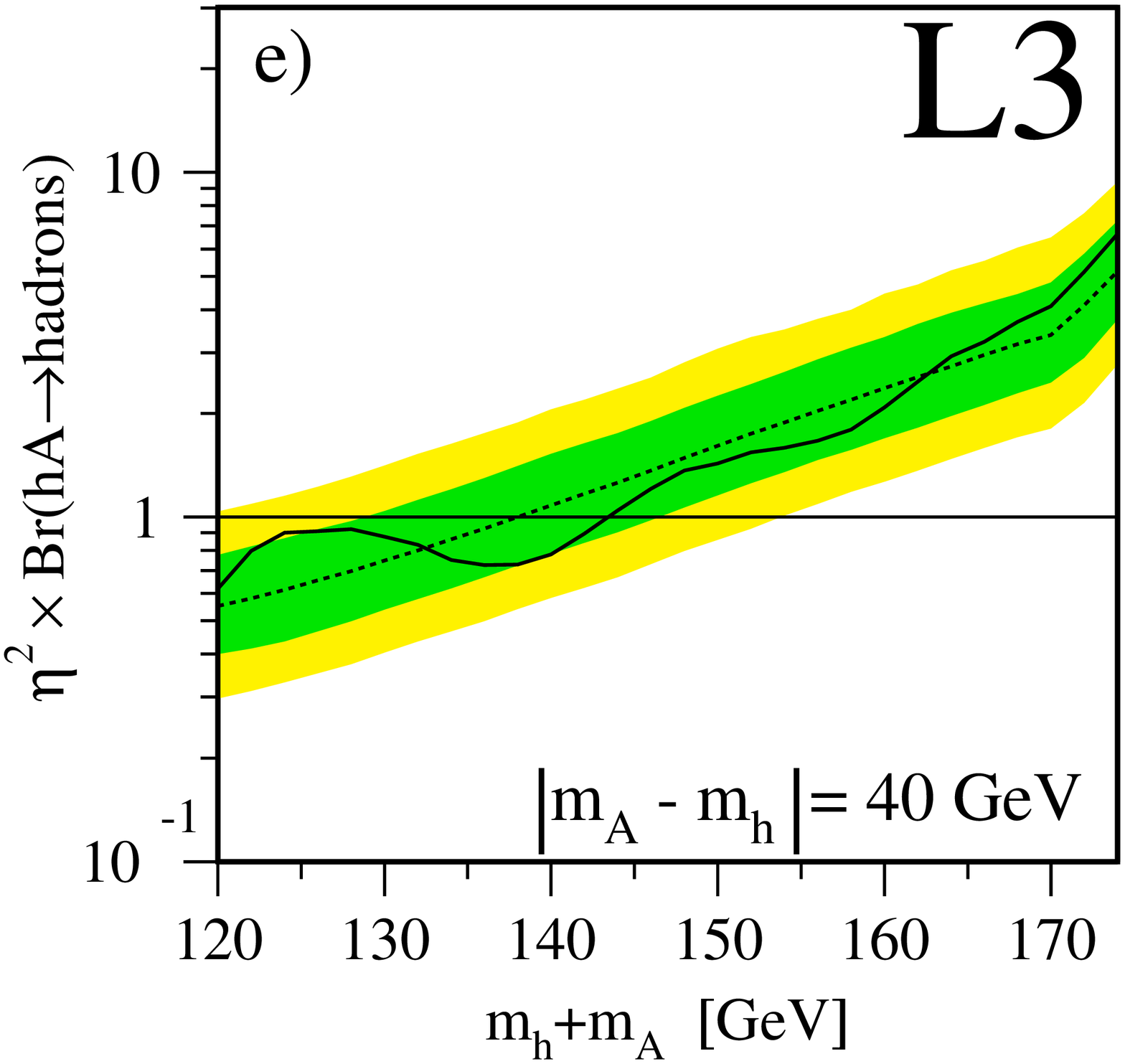} &
\hspace{-0mm}
\includegraphics*[width=0.4\textwidth]{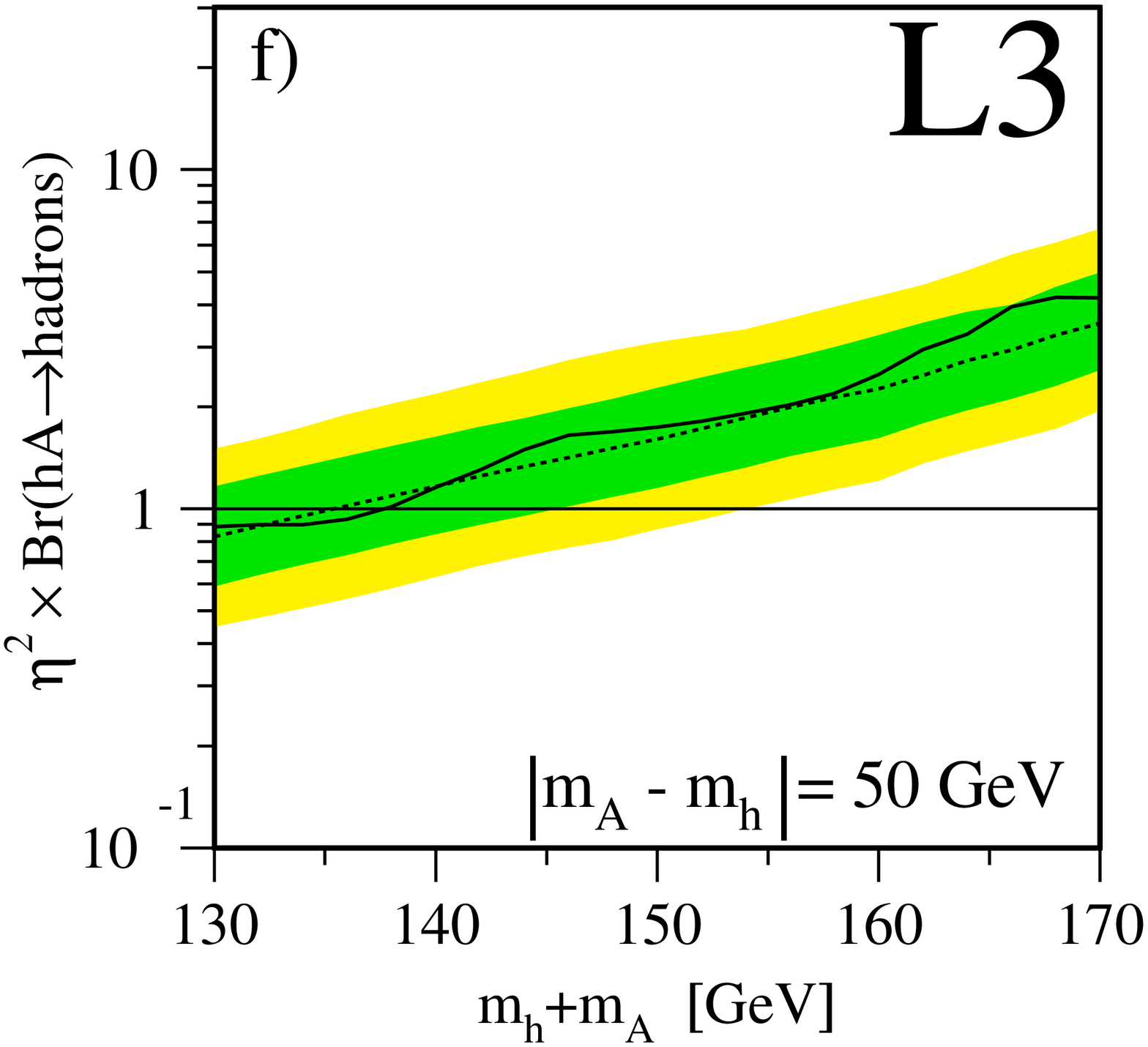} \\
\end{tabular}
        \caption[]{\label{fig:hA-xsec}
        The 95\% confidence level  upper limit on the 
        quantity $\mathrm{\eta^2\times Br(hA\ra hadrons)}$ as a function 
        of $\mh+\mA$ for different values of the
	difference: 
        $|\mA-\mh|$: 
	a) 0\GeV,
	b) 10\GeV,
	c) 20\GeV,
	d) 30\GeV,
	e) 40\GeV and
	f) 50\GeV.
        The solid lines indicate the
        observed limits and the dashed lines stand for the 
        median expected limits. The shaded areas show the $1\sigma$
        and $2\sigma$ intervals centered on the median expected limits.
        }
\end{center}
\end{figure}

Figure~\ref{fig:hA-xsec} shows the 95\% CL upper limit on
$\mathrm{\eta^2\times B(hA\ra hadrons)}$ as a function of $\mh+\mA$
for several values of $|\mh-\mA|$. The expected limits and the
$1\sigma$ and $2\sigma$ probability bands in absence of a signal are
also shown. The observed limits for $\eta=1$ are between 120 and
140$\GeV$, as expected. An excess of $2.9\sigma$ is observed arond $135\GeV$
for the
$\mh=\mA$ hypothesis. A similar behaviour is also observed in the
search for charged Higgs bosonse~\cite{chargedHiggs}. This excess is
mainly due to data at low values of $\sqrt{s}$. At higher energies and
for larger integrated luminosities it does not scale with the cross
section expected for a $\rm e^+e^-\rightarrow hA$ signal. It is hence
ascribed to a statistical fluctuation.

In conclusion, a flavour independent search for \h and \A bosons
produced through Higgs-strahlung or in pairs and decaying into
hadrons, shows no evidence of a signal and further constrains the
scenario of Higgs bosons light enough to have been produced at LEP.

%%%%%%%%%%%%%%%%%%%%%%%%%%%%%%%%%%%%%%%%%%%%%%%%%%%%%%%%%%%%%%%%%%%%%

\bibliographystyle{l3stylem}
\bibliography{flavourIndependent}

%%%%%%%%%%%%%%%%%%%%%%%%%%%%%%%%%%%%%%%%%%%%%%%%%%%%%%%%%%%%%%%%%%%%%

\newpage
\typeout{   }     
\typeout{Using author list for paper 279 -  }
\typeout{$Modified: Jul 15 2001 by smele $}
\typeout{!!!!  This should only be used with document option a4p!!!!}
\typeout{   }
%
%
%
%  L A T E X  version!!
%
%
% Make sure that the Lep package has been used!
%\input{Lep.sty}%
%
%\ifx\LepCalled\undefined%
%\typeout{     }%
%\typeout{!!!!!!!!!!!!!!!!!!!!!!!!!!!!!!!!!!!!!!!!!!!!!!!!!!!!!!!!!!!}%
%\typeout{Yikes.  You haven't used the Lep package!}%
%\typeout{Please put \protect\usepackage\protect{Lep\protect} in your preamble,
%         followed by}%
%\typeout{\protect\Lep\protect{1\protect} or \protect\Lep\protect{2\protect}}%
%\typeout{     }%
%\typeout{For now you will get a Lep phase 2 authorlist (may not be right!).}%
%\typeout{!!!!!!!!!!!!!!!!!!!!!!!!!!!!!!!!!!!!!!!!!!!!!!!!!!!!!!!!!!!}%
%\typeout{     }%
%\Lep{2}\fi%

\newcount\tutecount  \tutecount=0
\def\tutenum#1{\global\advance\tutecount by 1 \xdef#1{\the\tutecount}}
\def\tute#1{$^{#1}$}
\tutenum\aachen            % 1
\tutenum\nikhef            % 2
\tutenum\mich              % 3
\tutenum\lapp              % 4
\tutenum\basel             % 5
\tutenum\lsu               % 6
\tutenum\beijing           % 7
\tutenum\bologna           % 8
\tutenum\tata              % 9 
\tutenum\ne                % 10
\tutenum\bucharest         % 11
\tutenum\budapest          % 12
\tutenum\mit               % 13
\tutenum\panjab            % 14 
\tutenum\debrecen          % 15
\tutenum\dublin            % 16
\tutenum\florence          % 17
\tutenum\cern              % 18
\tutenum\wl                % 19
\tutenum\geneva            % 20
\tutenum\hefei             % 21
\tutenum\lausanne          % 22
\tutenum\lyon              % 23
\tutenum\madrid            % 24
\tutenum\florida           % 25
\tutenum\milan             % 26
\tutenum\moscow            % 27
\tutenum\naples            % 29
\tutenum\cyprus            % 30
\tutenum\nymegen           % 31
\tutenum\caltech           % 32
\tutenum\perugia           % 33
\tutenum\peters            % 34
\tutenum\cmu               % 35
\tutenum\potenza           % 36
\tutenum\prince            % 37
\tutenum\riverside         % 38
\tutenum\rome              % 39
\tutenum\salerno           % 40
\tutenum\ucsd              % 41
\tutenum\sofia             % 42
\tutenum\korea             % 43
\tutenum\purdue            % 44
\tutenum\psinst            % 45
\tutenum\zeuthen           % 46
\tutenum\eth               % 47
\tutenum\hamburg           % 48
\tutenum\taiwan            % 49
\tutenum\tsinghua          % 50

{
\parskip=0pt
\noindent
{\bf The L3 Collaboration:}
\ifx\selectfont\undefined%  old style font selection
 \baselineskip=10.8pt
 \baselineskip\baselinestretch\baselineskip
 \normalbaselineskip\baselineskip
 \ixpt
\else%                      new style font selection
 \fontsize{9}{10.8pt}\selectfont
\fi
\medskip
\tolerance=10000
\hbadness=5000
\raggedright
\hsize=162truemm\hoffset=0mm
\def\r{\rlap,}
\noindent

P.Achard\r\tute\geneva\ 
O.Adriani\r\tute{\florence}\ 
M.Aguilar-Benitez\r\tute\madrid\ 
J.Alcaraz\r\tute{\madrid}\ 
G.Alemanni\r\tute\lausanne\
J.Allaby\r\tute\cern\
A.Aloisio\r\tute\naples\ 
M.G.Alviggi\r\tute\naples\
H.Anderhub\r\tute\eth\ 
V.P.Andreev\r\tute{\lsu,\peters}\
% ...
\newpage
%\rule{\textwidth}{0.4pt}
\begin{list}{A}{\itemsep=0pt plus 0pt minus 0pt\parsep=0pt plus 0pt minus 0pt
                \topsep=0pt plus 0pt minus 0pt}
\item[\aachen]
 III. Physikalisches Institut, RWTH, D-52056 Aachen, Germany$^{\S}$
\item[\nikhef] National Institute for High Energy Physics, NIKHEF, 
     and University of Amsterdam, NL-1009 DB Amsterdam, The Netherlands
\item[\mich] University of Michigan, Ann Arbor, MI 48109, USA
\item[\lapp] Laboratoire d'Annecy-le-Vieux de Physique des Particules, 
     LAPP,IN2P3-CNRS, BP 110, F-74941 Annecy-le-Vieux CEDEX, France
\item[\basel] Institute of Physics, University of Basel, CH-4056 Basel,
     Switzerland
\item[\lsu] Louisiana State University, Baton Rouge, LA 70803, USA
\item[\beijing] Institute of High Energy Physics, IHEP, 
  100039 Beijing, China$^{\triangle}$ 
\item[\bologna] University of Bologna and INFN-Sezione di Bologna, 
     I-40126 Bologna, Italy
\item[\tata] Tata Institute of Fundamental Research, Mumbai (Bombay) 400 005, India
\item[\ne] Northeastern University, Boston, MA 02115, USA
\item[\bucharest] Institute of Atomic Physics and University of Bucharest,
     R-76900 Bucharest, Romania
\item[\budapest] Central Research Institute for Physics of the 
     Hungarian Academy of Sciences, H-1525 Budapest 114, Hungary$^{\ddag}$
\item[\mit] Massachusetts Institute of Technology, Cambridge, MA 02139, USA
\item[\panjab] Panjab University, Chandigarh 160 014, India.
\item[\debrecen] KLTE-ATOMKI, H-4010 Debrecen, Hungary$^\P$
\item[\dublin] Department of Experimental Physics,
  University College Dublin, Belfield, Dublin 4, Ireland
\item[\florence] INFN Sezione di Firenze and University of Florence, 
     I-50125 Florence, Italy
\item[\cern] European Laboratory for Particle Physics, CERN, 
     CH-1211 Geneva 23, Switzerland
\item[\wl] World Laboratory, FBLJA  Project, CH-1211 Geneva 23, Switzerland
\item[\geneva] University of Geneva, CH-1211 Geneva 4, Switzerland
\item[\hefei] Chinese University of Science and Technology, USTC,
      Hefei, Anhui 230 029, China$^{\triangle}$
\item[\lausanne] University of Lausanne, CH-1015 Lausanne, Switzerland
\item[\lyon] Institut de Physique Nucl\'eaire de Lyon, 
     IN2P3-CNRS,Universit\'e Claude Bernard, 
     F-69622 Villeurbanne, France
\item[\madrid] Centro de Investigaciones Energ{\'e}ticas, 
     Medioambientales y Tecnol\'ogicas, CIEMAT, E-28040 Madrid,
     Spain${\flat}$ 
\item[\florida] Florida Institute of Technology, Melbourne, FL 32901, USA
\item[\milan] INFN-Sezione di Milano, I-20133 Milan, Italy
\item[\moscow] Institute of Theoretical and Experimental Physics, ITEP, 
     Moscow, Russia
\item[\naples] INFN-Sezione di Napoli and University of Naples, 
     I-80125 Naples, Italy
\item[\cyprus] Department of Physics, University of Cyprus,
     Nicosia, Cyprus
\item[\nymegen] University of Nijmegen and NIKHEF, 
     NL-6525 ED Nijmegen, The Netherlands
\item[\caltech] California Institute of Technology, Pasadena, CA 91125, USA
\item[\perugia] INFN-Sezione di Perugia and Universit\`a Degli 
     Studi di Perugia, I-06100 Perugia, Italy   
\item[\peters] Nuclear Physics Institute, St. Petersburg, Russia
\item[\cmu] Carnegie Mellon University, Pittsburgh, PA 15213, USA
\item[\potenza] INFN-Sezione di Napoli and University of Potenza, 
     I-85100 Potenza, Italy
\item[\prince] Princeton University, Princeton, NJ 08544, USA
\item[\riverside] University of Californa, Riverside, CA 92521, USA
\item[\rome] INFN-Sezione di Roma and University of Rome, ``La Sapienza",
     I-00185 Rome, Italy
\item[\salerno] University and INFN, Salerno, I-84100 Salerno, Italy
\item[\ucsd] University of California, San Diego, CA 92093, USA
\item[\sofia] Bulgarian Academy of Sciences, Central Lab.~of 
     Mechatronics and Instrumentation, BU-1113 Sofia, Bulgaria
\item[\korea]  The Center for High Energy Physics, 
     Kyungpook National University, 702-701 Taegu, Republic of Korea
\item[\purdue] Purdue University, West Lafayette, IN 47907, USA
\item[\psinst] Paul Scherrer Institut, PSI, CH-5232 Villigen, Switzerland
\item[\zeuthen] DESY, D-15738 Zeuthen, Germany
\item[\eth] Eidgen\"ossische Technische Hochschule, ETH Z\"urich,
     CH-8093 Z\"urich, Switzerland
\item[\hamburg] University of Hamburg, D-22761 Hamburg, Germany
\item[\taiwan] National Central University, Chung-Li, Taiwan, China
\item[\tsinghua] Department of Physics, National Tsing Hua University,
      Taiwan, China
\item[\S]  Supported by the German Bundesministerium 
        f\"ur Bildung, Wissenschaft, Forschung und Technologie
\item[\ddag] Supported by the Hungarian OTKA fund under contract
numbers T019181, F023259 and T037350.
\item[\P] Also supported by the Hungarian OTKA fund under contract
  number T026178.
\item[$\flat$] Supported also by the Comisi\'on Interministerial de Ciencia y 
        Tecnolog{\'\i}a.
\item[$\sharp$] Also supported by CONICET and Universidad Nacional de La Plata,
        CC 67, 1900 La Plata, Argentina.
\item[$\triangle$] Supported by the National Natural Science
  Foundation of China.
\end{list}
}
\vfill

%%% Local Variables: 
%%% mode: latex
%%% TeX-master: t
%%% End:

\newpage

%%%%%%%%%%%%%%%%%%%%%%%
%       TABLES
%%%%%%%%%%%%%%%%%%%%%%%

\begin{table}
\begin{center}
\begin{tabular}{|r|cccccccccc|}
\hline
$\sqrt{s}$ (GeV) & 
 188.6 & 
 191.6 & 
 195.6 & 
 199.5 & 
 201.5 & 
 203.8 & 
 205.1 & 
 206.3 & 
 206.6 & 
 208.0 \\ 
$\cal{L}$ (pb$^{-1})$ & 
176.4            &
\phantom{0}29.7 &
\phantom{0}83.7 &
\phantom{0}82.8 &
\phantom{0}37.0 &
\phantom{00}7.6 &
\phantom{0}68.1 &
\phantom{0}66.9 &
\phantom{0}63.7 &
\phantom{00}8.2 \\

\hline
\end{tabular}
\end{center}
        \caption[]{\label{tab:data}
        Effective centre-of-mass energies and 
        corresponding integrated luminosities, $\cal{L}$.}
\end{table}

\begin{table}
  \begin{center}
    \begin{tabular}{|c|c|c|c|}
      \hline
      \multicolumn{2}{|c|}{Process} & \multicolumn{2}{c|}{Process}\\
      \hline
      \multicolumn{2}{|c|}{$\rm \epem\ra\h Z$} & \multicolumn{2}{c|}{$\epem\ra\h\A$}\\
      \multicolumn{2}{|c|}{$\rm\h\ra b\bar{b}, c\bar{c},  gg$\,\,\,\, $\rm Z\ra q\bar{q}, \nu\bar{\nu}, \ell^+\ell^-$} & \multicolumn{2}{c|}{$\rm\h\ra b\bar{b}, c\bar{c},  gg$\,\,\,\, $\rm\A\ra b\bar{b}, c\bar{c},  gg$}\\
      \hline
      Final state & Topology & Final State & Topology \\
      \hline
      \rule{0pt}{12pt}$\rm  b\bar{b} q\bar{q}$, $\rm  c\bar{c} q\bar{q}$, $\rm  gg q\bar{q}$ &
      Four jets &
      $\rm  b\bar{b} b\bar{b}$,  $\rm  b\bar{b} c\bar{c}$ & \\
      \rule{0pt}{12pt}$\rm  b\bar{b} \nu\bar{\nu}$, $\rm  c\bar{c} \nu\bar{\nu} $, $\rm  gg
      \nu\bar{\nu} $ &
      Two jets and missing energy &
      $\rm  b\bar{b} gg$,  $\rm  c\bar{c} c\bar{c}$ & Four jets\\
      \rule{0pt}{12pt}$\rm  b\bar{b} \ell^+\ell^-$, $\rm  c\bar{c} \ell^+\ell^- $, $\rm 
      gg\ell^+\ell^-$ &
      Two jets and two leptons &
      $\rm  c\bar{c} gg$, $\rm gggg$ & \\
      \hline
    \end{tabular}
  \end{center}
  \caption[]{\label{tab:signatures}
   Final states of the $\rm e^+e^-\rightarrow hZ$ and  $\rm
   e^+e^-\rightarrow hA$ processes and topologies under study.}
\end{table}

\begin{table}
\begin{center}
\begin{tabular}{|c|cccc|cccc|}
\hline
\multicolumn{9}{|c|}{$\rm e^+e^-\rightarrow hZ$} \\ 
\hline       
 &  
\multicolumn{4}{c|}{$\mathrm{h\rightarrow hadrons}$ }    &    
\multicolumn{4}{c|}{$\mathrm{h\rightarrow hadrons}$ }    \\   
 &  
\multicolumn{4}{c|}{$\mathrm{Z\rightarrow q\bar{q}}$}    &    
\multicolumn{4}{c|}{$\mathrm{Z\rightarrow\nu\bar{\nu}}$} \\
$\mh$  (GeV)     & $N_{\rm D}$ & $N_{\rm B}$ & $N_{\rm S}$ & $\varepsilon$ (\%) &
                   $N_{\rm D}$ & $N_{\rm B}$ & $N_{\rm S}$ & $\varepsilon$ (\%) \\
\hline
\phantom{2}60              & 
1356            &   1336        &  172\phantom{.0}       & 43 &
\phantom{2}40   &   \phantom{1}33.1        &  48.8                 &  63 \\
\phantom{2}70              & 
1363            &   1295        &  122\phantom{.0}       & 43 &
\phantom{2}89   &   \phantom{1}84.0        &  40.6                 &  60 \\
\phantom{2}80              &
\phantom{1}938  &\phantom{1}966 &  104\phantom{.0}       & 45 &
209             &201\phantom{.0} &  32.2                 &  56 \\
\phantom{2}90              &
\phantom{1}584  &\phantom{1}585 &  \phantom{0}71.2                 & 45 &
183             &181\phantom{.0} &  21.6                 & 53 \\
100             &
\phantom{1}360  &\phantom{1}355 &  \phantom{0}39.9                 & 46 &
\phantom{1}74   & \phantom{1}69.9          &  12.2                 & 50 \\
110             &
\phantom{1}126  &\phantom{1}127 &  \phantom{0}11.8                 & 46 &
\phantom{1}18   & \phantom{1}16.4          &  \phantom{1}3.5       & 48 \\
\hline
&
\multicolumn{4}{c|}{$\mathrm{h\rightarrow hadrons}$ }    & 
\multicolumn{4}{c|}{}                                    \\ 
&  
\multicolumn{4}{c|}{$\mathrm{Z\rightarrow\ell^+\ell^-}$}  & 
\multicolumn{4}{c|}{Combined}                            \\ 
$\mh$  (GeV)     & $N_{\rm D}$ & $N_{\rm B}$ & $N_{\rm S}$ & $\varepsilon$ (\%) &
                   $N_{\rm D}$ & $N_{\rm B}$ & $N_{\rm S}$ & \\
\hline
\phantom{2}60              & 
  49            &   49.4        &  28.2                 &  49 &
1445            &   1419        &  249\phantom{.0}      & \\  
\phantom{2}70              & 
43              &   52.8        &  24.1                 &  50 &
1495            &   1432        &  187\phantom{.0}      & \\
\phantom{2}80              &
61              &  63.2         &  19.2                 &  51 &
1208            &  1230         &  155\phantom{.0}      & \\      
\phantom{2}90              &
 56             & 61.3          &  13.0                 & 50 &
\phantom{1}823  &\phantom{1}827 &  106\phantom{.0}      & \\
100             &
 24             & 18.4          &  \phantom{1}5.8       & 47 &
\phantom{1}458  &\phantom{1}443 &  \phantom{0}57.9                 &\\
110             &
\phantom{2}3    &\phantom{1}4.2 &  \phantom{1}1.6       & 42 &
\phantom{1}147  &\phantom{1}148 &  \phantom{0}16.9                 &\\
\hline
\end{tabular}
\end{center}
        \caption[]{\label{tab:hz}
	Numbers of selected candidates, $N_D$, expected background
        events, $N_B$, and expected signal events, $N_S$, for
        different $m_{\rm h}$ hypotheses in the $\rm e^+e^-\rightarrow
        hZ$ search. The selection efficiencies, $\varepsilon$, are also given.
	The numbers of signal events are quoted for the h
        decay mode corresponding to the lowest efficiency and are
        computed assuming $\mathrm{\xi^2\times B(h\ra hadrons)}$ =
	1. Only events with a signal over background ratio greater than
	0.05 are considered.}
\end{table}

%%%%%%%%%%%%%%%%%%
\begin{table}
\begin{center}
\begin{tabular}{|c|cccc|}
\hline
\multicolumn{5}{|c|}{$\rm e^+e^-\rightarrow hA$}                \\
\hline
($\mh$,$\mA$) (GeV)                   &  $N_D$        &  $N_B$        &  $N_S$  &    $\varepsilon$ (\%)   \\
\hline
(50,50)                 &  114          &110\phantom{.0} & 84.4          & 41\\
(50,70)                 &  220          &211\phantom{.0} & 56.2          & 36\\
(50,90)                 &  223          &239\phantom{.0} & 29.2          & 28\\
(70,70)                 &  244          &223\phantom{.0} & 39.9          & 40\\
(70,90)                 &\phantom{2}96  &  95.8          &\phantom{3}5.7 & 11\\
(90,90)                 &\phantom{2}10  &  11.4          &\phantom{3}0.6            & \phantom{0}4\\
\hline
\end{tabular}
\end{center}
        \caption[]{\label{tab:ha}
	Number of selected candidates, $N_D$, expected background
        events, $N_B$, and expected signal events, $N_S$, and selection efficiencies, $\varepsilon$, for
        different $(m_{\rm h},m_{\rm A})$ hypotheses in the $\rm e^+e^-\rightarrow
        hA$ search. The numbers of signal events are quoted for the h
	and A
        decay modes corresponding to the lowest efficiencies and are
        computed assuming $\mathrm{\eta^2\times B(h\ra hadrons)}$ =
	1. Only events with a signal over background ratio greater than
	0.05 are considered.}
\end{table}

\newpage

\end{document}